\documentclass[aip,showpacs,nofootinbib,floatfix,superscriptaddress,pre,11pt]{revtex4-1}
\usepackage{amsmath}
\usepackage{amsfonts}
\usepackage{tikz}
\usetikzlibrary{snakes}
\usepackage{amsmath}
\usepackage{amssymb}
\usepackage{CJKutf8}
\usepackage{verbatim,graphicx}
\usepackage{mathtools}
\usepackage{color}
\usepackage[all]{xy}
\usepackage{simplewick}
\usepackage{hyperref}
\usepackage{setspace}
\usepackage{array}
\usepackage{tikz}
\usepackage{tkz-euclide}
\usepackage{amsthm}
\usepackage{enumitem}
\usepackage{bbm}
\usepackage{bbold}
\usepackage{hyperref}

\graphicspath{{figs-q2/}}

\newcounter{jvcc}
\newcounter{dario}

\newcounter{yjcc}
\newcommand{\mj}{\mathcal J}
\newcommand{\mjt}{\widetilde{\mathcal J}}

\newcommand{\Tr}{{\rm Tr}}

\newcommand{\eref}[1]{(\ref{#1})}
\newcommand{\nn}{\nonumber}

\newcommand{\sign}{\rm sign}
\newcommand{\be}{\begin{eqnarray}}
\newcommand{\ee}{\end{eqnarray}}
\newcommand{\vev}[1]{\langle #1\rangle}

\newcommand{\elambda}{\varepsilon}

\newcommand{\vect}{\left ( \begin{array}{c} }
	\newcommand{\evect}{\end{array} \right ) }
\newcommand{\bmat}{\left ( \begin{array}{cc} }
	\newcommand{\emat}{\end{array} \right ) }

\usetikzlibrary{shapes.misc}

\tikzset{cross/.style={cross out, draw=black, fill=none, minimum size=2*(#1-\pgflinewidth), inner sep=0pt, outer sep=0pt}, cross/.default={2pt}}

\def\Tr{\textrm{Tr}}

\usetikzlibrary{shapes.misc}
  \newcommand{\E}{\epsilon}
  \newcommand{\y}{\upsilon}

\newcommand{\beq}{\begin{equation}}
\newcommand{\beqs}{\begin{equation*}}
\newcommand{\eeq}{\end{equation}}
\newcommand{\eeqs}{\end{equation*}}

\allowdisplaybreaks

\begin{document}

\title{Replica Symmetry Breaking  for the
  Integrable Two-Site Sachdev-Ye-Kitaev Model}
\author{Yiyang Jia\begin{CJK*}{UTF8}{gbsn}
(贾抑扬)
\end{CJK*}}
\email{yiyang.jia@weizmann.ac.il}
\affiliation{Department of Particle Physics and Astrophysics, Weizmann Institute of Science, Rehovot 7610001, Israel}
\author{Dario Rosa}
\email{dario\_rosa@ibs.re.kr}
\affiliation{Center for Theoretical Physics of Complex Systems,
  Institute for Basic Science (IBS), Daejeon - 34126, Korea}
\author{Jacobus J. M. Verbaarschot}
\email{jacobus.verbaarschot@stonybrook.edu}
\affiliation{Center for Nuclear Theory, Department of Physics and Astronomy, Stony Brook University, Stony Brook, New York 11794, USA}
\begin{abstract}
  We analyze a two-body nonhermitian two-site Sachdev-Ye-Kitaev model
  with the couplings of one site complex conjugated to the other site.
  This model, with no explicit coupling between the sites, shows an
  infinite number of second order phase transitions which is a consequence of the
  factorization of the partition function into a product over Matsubara frequencies.
  We calculate the quenched free energy in two different ways, first in terms
  of the single-particle energies, and second by solving the Schwinger-Dyson
  equations of the two-site model.
  The first calculation can be done entirely
  in terms of a one-site model. The conjugate replica enters due to non-analyticities when Matsubara frequencies enter the spectral support of the coupling
  matrix. The second calculation is based on the replica trick of the two-site partition function. Both methods give the same result.
  The free-fermion partition function can also be rephrased as
  a matrix model for the coupling matrix. Up to minor details, this model
  is the random matrix model that describes the chiral phase transition
  of QCD,  and the order parameter of the two-body model corresponds to the
  chiral condensate of
  QCD. Comparing to the corresponding four-body model,
we are able to determine which features of the free energy are
due to chaotic nature of the four-body model. The high-temperature phase of
both models is entropy dominated, and in both cases the free energy is
determined by the spectral density. The chaotic four-body SYK model has a low-temperature phase whose free energy is almost temperature-independent,  signaling an effective gap of the theory even though the actual spectrum does not exhibit a gap.  On the other hand the low-temperature free energy of the two-body SYK model is not flat,  in fact it oscillates to arbitrarily low temperature.  This indicates a less
desirable feature  that the entropy of the two-body model is  not
always positive in the low-temperature phase, which most likely is a
consequence of the nonhermiticity.
\end{abstract}

\maketitle
\tableofcontents
\section{Introduction}

In 1971 George Uhlenbeck asked Freeman Dyson ``Which nucleus has levels
distributed according to the semi-circle law?'' \cite{dyson1996selected}.
As answer to his criticism, Dyson published his last paper on Random Matrix
Theory \cite{Dyson:1972tm}
in which he introduced a Brownian motion process to construct an
ensemble of random matrices with an arbitrary level density, in particular
the nuclear level density given by the Bethe formula \cite{bethe1936},
\be
\rho(E) =b e^{c\sqrt E}.
\ee
Uhlenbeck could have asked a different question, namely ``Is the nuclear force
an all--to--all many-body interaction?'' which is the case for the Wigner-Dyson
ensembles.
This question led French, Wong, Bohigas,
Flores and Mon
\cite{french1970,french1971,bohigas1971,bohigas1971a,mon1975}
around the same time to the introduction of the two-body (which in the SYK
literature is known as a four-body interaction)
random ensemble which
reflects the two-body nature of the nuclear interaction. It took until
2015, through the seminal work of Kitaev \cite{kitaev2015,maldacena2016},
to realize that  Uhlenbeck's criticism could also have
 been addressed by this work. The two-body random ensemble,
in particular the version of the model introduced by Mon and French \cite{mon1975},
is now known as the complex Sachdev-Ye-Kitaev (SYK) model \cite{sachdev1993,sachdev2015}.

Since the pioneering work of Wigner, Dyson, Gaudin and Mehta \cite{wigner1951,mehta1960density,dyson1962statistical,dyson1962statisticalII,dyson1962statisticalIII,dyson1963statisticalI,mehta1963statisticalII,dyson1962threefold,Dyson:1972tm} random matrix theory
has been applied to virtually all areas of physics and even outside of physics,
see the  comprehensive review by Guhr, M\"uller Groeling and Weidenm\"uller \cite{guhr1998}  . In this paper we study
nonhermitian random matrix theories which were first introduced by Ginibre
\cite{ginibre1965}, but have also been applied to many areas of physics.
For example, the distribution of poles of S-matrices \cite{Verbaarschot1984a,Verbaarschot:1985jn,sommers1999s}, the Hatano-Nelson model
\cite{hatano1996localization,efetov1997directed,brouwer1998delocalization}, dissipative quantum systems
\cite{fyodorov1997,akemann2019universal,li2021spectral,sa2021lindbladian},
QCD at nonzero chemical potential \cite{stephanov1996,Janik:1996va,Verbaarschot:2000dy,Osborn:2004rf,Akemann:2004dr,Osborn:2005ss,Kanazawa:2009en,kanazawa2021new,kanazawa2021complex} and PT-symmetric systems \cite{bender1998}, to mention a few.
Nonhermitian random matrices were classified 
\cite{halasz:1997fc,bernard2002classification,Magnea:2007yk,kawabata2019symmetry,Garcia-Garcia:2021rle}
along the lines of the classification
of Hermitian random matrices \cite{dyson1962threefold,Dyson:1972tm,Verbaarschot:1994qf,Altland:1997zz}. A recent review of nonhermitian physics was given by
Ashida, Gong and Ueda \cite{ashida2020non}.

The possibility  of a  nonhermitian version of the Sachdev-Ye-Kitaev (SYK) model
was originally suggested
by Maldacena and Qi \cite{maldacena2018,Garcia-Garcia:2019poj}
as a two-site SYK model 
for Euclidean wormholes without explicit coupling between the two
SYK models (the only coupling is through the randomness) and was studied
in detail in subsequent papers \cite{Garcia-Garcia:2020ttf,Garcia-Garcia:2021elz,Garcia-Garcia:2021rle,Garcia-Garcia:2022}. In this model,  the Left (L) and Right (R) partition functions
are complex conjugate to each other,  and each of them has $N/2$ Majorana fermions so that
energy levels of the $q$-body Hamiltonian
\be
H = H_L \otimes \mathbb{1} + \mathbb{1} \otimes H_R,
\ee
are given $\{E_k + E_l^*\}$ if the $\{E_k\}$ are the eigenvalues of $H_L$.
 Therefore the partition function  factorizes as
\be
Z = Z_L Z_R = Z_L Z_L^*
\ee
and is necessarily positive definite.

One of the main conclusions from these studies is that
the chaotic  model has a first order phase transition which separates the low-temperature
phase from the high-temperature phase. In the high-temperature phase,
the average of the partition function factorizes
\be
\langle Z\rangle = \langle Z_L \rangle   \langle Z_L^* \rangle
\ee
and
the free energy follows from the eigenvalue density of the one-site Hamiltonian.
Because of the complex phases of the eigenvalues the average partition
function is exponentially suppressed due to cancellations. In the case of maximum nonhermiticity,
when the eigenvalue density is isotropic in the complex plane, the partition
function $ \langle Z_L \rangle$ becomes temperature independent and the
free energy is determined by the total number of states $F = -T \frac N2 \log 2$.
It is clear that this result cannot be correct at low temperature when
$F \to -|E_0|$ with $E_0$ the ground state energy. The only possibility
is that the low-temperature phase receives contributions from the
correlations of the eigenvalues of the $L$ and $R$ Hamiltonians. Indeed,
for the $q=4$ SYK model the free energy in the low-temperature phase
is entirely determined by the two-point correlations of the
eigenvalues \cite{Garcia-Garcia:2021elz,Garcia-Garcia:2022}.
The reason that this can happen is the exponential suppression of the
single site partition function due to the complex phase of the eigenvalues.
The dynamics of the $q=4$ SYK model is chaotic with eigenvalue correlations
in the universality class of the  Ginibre model.   The universal two-point correlations give rise to a temperature-independent
free energy at low temperatures.
This also explains that results obtained by
solving the Schwinger-Dyson equations are very close to the results
for the Ginibre  random matrix ensemble.
We conclude that the quantum chaotic nature
of the model is responsible for a nearly temperature-independent\footnote{There
  are small deviations from $-E_0$ when the temperature becomes closer
  to $T_c$. The nature of these deviations is not clear.}
free energy in the low-temperature phase when an actual spectral gap is absent.
As a consequence, the free energy
of the low-temperature phase of the $q=2$ SYK model, which is integrable and has no spectral gap, has
to be different. The goal of  this paper is to solve the $q=2$ nonhermitian SYK
model to study the effects of chaos and integrability on the phase diagram.

As was the case for the $q=4$ SYK model \cite{Garcia-Garcia:2021elz,Garcia-Garcia:2022}, in this paper we evaluate the {\it quenched} free energy in two structurally  different ways.
First, from the eigenvalues
of the SYK Hamiltonian giving the quenched free energy,
and second, from the solution of the Schwinger-Dyson
equations \cite{dyson1949s,schwinger1951green} 
of the SYK model
in $ \Sigma G$ formulation, giving the {\it annealed } free energy of the
two-site SYK model. 
For $q=2$ it is possible to perform the spectral calculation both analytically
and numerically, while for $q=4$ this could only be done numerically by
an explicit diagonalization of the SYK Hamiltonian. Also the Schwinger-Dyson equation can be solved analytically for $q=2$ \cite{Maldacena:2015waa,Cotler:2016fpe} while
for $q=4$ we had to rely on numerical techniques.

The  $\Sigma G$ formulation of the SYK model is based on the replica
trick for the quenched free energy \cite{edwards1971statistical}
\be
\log Z = \lim_{n\to 0} \frac {Z^n-1}{n},
\ee
which is known to fail \cite{sherrington1972,  verbaarschot1985, zirnbauer1999critique} in particular
for nonhermitian theories \cite{Barbour:1986jf}. However, by now it has
been well understood how to refine the replica method so that its 
results can be trusted \cite{parisi1979,girko2012theory,stephanov1996,mezard1999,nishigaki2002a,kanzieper2002,splittorff:2003cu,sedrakyan2005toda}. For Hermitian models the naive replica
trick usually gives the
correct result for a mean field analysis. This is also the case for
the $\Sigma G$ formulation of the SYK model \cite{arefeva2018,wang2018}.
However, as should be clear from the arguments given above,
the naive application of the replica trick to
the one-site nonhermitian SYK model gives an incorrect result for the
quenched free energy of the
low-temperature phase. Instead, the quenched free energy
of the one-site Hamiltonian
is given by the replica limit of the product of the partition function
and its complex conjugate. In the case of the one-site partition function,
the replica symmetry is broken in the sense that the
conjugate replica emerges due to quenching and couples to the original
replica.
The emergence of conjugate replicas for quenched replicas
is well-known in nonhermitian
RMT \cite{girko2012theory,efetov1997directed,feinberg1997non} and QCD at nonzero chemical potential
\cite{stephanov1996,Janik:1996va,Janik:1996xm,splittorff:2003cu}.

We start this paper with a short introduction of the SYK model and the
replica trick. Since the $q=2$ SYK model is a Fermi liquid, we can calculate
the quenched free energy from a free-fermion formulation of the model, see
section III. In section IV we calculate annealed free energy for the $\Sigma G$
formulation of the SYK model for one replica and one conjugate replica. The
final result is in complete agreement with the result from the free-fermion
calculation. The $\Sigma G$ action of the $q=2$ SYK model is quadratic in $G$
so that it can be integrated out exactly. The result resembles a random
matrix $\sigma$-model. In section V, we show that this $\sigma$-model can also
directly obtained from the free-fermion description of the $q=2$ SYK model.
Concluding remarks are made in section VI and some technical details are
worked out in two appendices.
In Appendix A we show that the
occupation number representation also applies to the nonhermitian SYK model,
and in Appendix B, we work out
 the free-fermion calculation for the case  the nonhermiticity is not maximal.

\section{\boldmath The $q=2$  nonhermitian SYK model}

     The Hamiltonian of the $q=2$ nonhermitian SYK model is given by
 \begin{align}
 \label{eq:sykq2}
  H &= i\sum_{i < j}^{N / 2} \left( J_{ij} + i \, k M_{ij} \right) \psi_L^i \psi_L^j - i\sum_{i < j}^{N / 2} \left( J_{ij} - i \, k M_{ij} \right) \psi_R^i \psi_R^j
\nn\\ &\equiv H_L\otimes \mathbb{1} +\mathbb{1} \otimes H_R,
 \end{align}
where  $\psi$ are Majorana fermions and $J, M$ are random couplings. A general $q$-body Hamiltonian would have products of $q$ Majorana fermions.  The tensor product structure on the second line of equation \eqref{eq:sykq2} follows from an explicit Dirac-matrix representation of $\psi$:
\begin{equation}\label{eqn:majoranaNormalization}
\begin{split}
\psi_L^i &= \frac{1}{\sqrt{2}} \gamma_i \otimes \mathbb{1}, \\
\psi_R^i &= \frac{1}{\sqrt{2}} \gamma_c \otimes \gamma_i,
\end{split}
\end{equation} 
where $\gamma_i$ are the Dirac matrices in $N/2$ dimensions and $\gamma_c$ is the corresponding chirality matrix (assuming $N/2$ is even).  The variances of the couplings are given by
\begin{equation}\label{eqn:variancePhysical}
\langle J_{ij}^2\rangle = \langle M_{ij}^2\rangle = {v^2}/{(N/2)}
\end{equation}
 where $v$ is a dimensionful parameter that sets the physical scale. 
 In a representation where the left gamma matrices are real and the right
 gamma matrices are purely imaginary, the Hamiltonian of a single SYK
 is anti-symmetric under transposition. The eigenvalues, which are
 representation independent thus occur in pairs $\pm \lambda_k$.  In this
 representation we also have that $H_L = H_{R}^*$ (with the minus sign
  from the sum included in $H_R$). The spectrum of the Hamiltonian
 \eref{eq:sykq2} is  given by $ \pm \lambda_k\pm \lambda_l^*$ if the $\lambda_k$
 are the eigenvalues of $H_L$ with a positive real part. The partition function
 of this Hamiltonian is necessarily positive
 \be
 Z  = Z_L Z_R = |Z_L|^2,
 \ee
 where $Z_{L(R)}$ is the partition function of the $L(R)$ Hamiltonian. 
 The average partition function will be denoted $\langle Z \rangle$ with the
 appropriate subscripts.
Contrary to $q>2$ the SYK Hamiltonian for $q=2$ is a Fermi liquid with
single particle energies $\pm \varepsilon_k$ given by the eigenvalues of the coupling matrix,
which is an anti-symmetric nonhermitian random matrix. Therefore the
quenched partition function is given by
\be
\langle \log Z\rangle &=& \left \langle \log \left[\prod_k (e^{-\beta \varepsilon_k}+e^{\beta \varepsilon_k})
\prod_k (e^{-\beta \varepsilon_k^*}+e^{\beta \varepsilon_k^*})\right ]\right \rangle\nn\\
&=&\sum_k \left \langle \log  (e^{-\beta \varepsilon_k}+e^{\beta \varepsilon_k})\right\rangle + \sum_k \left\langle 
\log (e^{-\beta \varepsilon_k^*}+e^{\beta \varepsilon_k^*})\right \rangle,
\ee
and
\be
\langle \log Z\rangle= \langle \log Z_L\rangle + \langle \log Z_R\rangle
= 2\langle \log Z_L\rangle.
\ee
For the last equality we have used that the average density of the single
particle energies satisfies
$\langle \rho_L^{\rm sp}(z)\rangle
=\langle \rho_R^{\rm sp}(z^*)\rangle$.
 Therefore, the quenched free energy can be obtained from the
one-site partition function, which is a direct consequence of the Fermi-liquid
nature of the $q=2$ SYK model.

When we evaluate the quenched free energy of the SYK partition function
in the $\Sigma G$ formulation, we will employ 
the replica  trick
 \be
 -\beta F = \langle \log Z \rangle = \lim_{n\to 0} \left \langle
 \frac {Z^n - 1} n \right \rangle.
 \ee
 For large $N$, the partition function can be evaluated by a saddle point
 approximation.
 As was argued in \cite{Garcia-Garcia:2021elz,Garcia-Garcia:2022},
 since the partition function $Z = Z_LZ_L^*$ is positive definite,
  the replica
trick is expected to give the correct mean field result with unbroken replica symmetry,
\be
\langle Z^n\rangle = \langle Z \rangle^n.
\ee
Therefore, the quenched free energy is equal to
the annealed free energy
\be
\langle \log Z \rangle = \log \langle Z \rangle,
\ee
and it 
can be calculated by evaluating the partition function for one replica of the two-site Hamiltonian,  in other words
one replica and one conjugate replica in terms of the one-site Hamiltonian. 
If the (left-right) replica symmetry is broken, we have that
\be
\langle (Z_L  Z^*_L)^n \rangle =\langle Z_L  Z^*_L \rangle^n  \ne \langle Z_L \rangle^n \langle Z^*_L \rangle^n
\ee
so that it is not guaranteed that the quenched free energy can be
obtained from the one-site annealed free energy. We will see in section
\ref{sec:sd} that is the case for the low-temperature phase of the
$\Sigma G$ formulation of the SYK model. In the literature, the coupling
between replicas has been
related to the formation of wormholes between black holes
\cite{Saad:2019lba,Altland:2021rqn}.

Before calculating the annealed free energy from the $\Sigma G$ formulation
of the SYK model, 
in the next section, we will evaluate the average partition function, using  
the properties of the spectra of anti-symmetric nonhermitian random matrices.

\section{\boldmath  The free energy of the two-site non-hermitian SYK for $q=2$}\label{sec:sykint}

Results for the $q=4$ nonhermitian SYK model \cite{Garcia-Garcia:2021elz,Garcia-Garcia:2022} suggest that quantum chaotic dynamics is responsible for a replica-symmetry-breaking (RSB) phase at low temperatures with a temperature-independent
free energy. Indeed, the free energy of  an integrable nonhermitian model of random uncorrelated energies \cite{Garcia-Garcia:2022} exhibits a distinct low-temperature phase with a temperature-dependent free energy.  However, the random energy model lacks a natural interpretation
as a many-body model. In this section, we test this hypothesis by evaluating quenched free
energy of the
two-site $q = 2$ non-Hermitian SYK model which is a Fermi liquid with energies given by sums of
     single particle energies. Therefore the many-body eigenvalues obey Poisson statistics
     which is consistent with a vanishing Lyapunov exponent \cite{garcia2018chaotic}
     (obtained by  calculating the Out of Time Order Correlator).
     We will calculate the quenched free energy from the single-particle energy  density which is
     constant inside an ellipse in the complex plane. As argued in previous
     section, the quenched free energy can be obtained from the {\it one-site} partition
     function. In section \ref{sec:sd} we will see that the same result
     can be obtained from the SD equation of the {\it two-site} model.

  After a change of basis (see appendix \ref{a:yiyangfreefermions} for a demonstration),  the one-site Hamiltonian can be expressed as a free Fermi liquid with single-particle
energies which are the eigenvalues of the antisymmetric coupling matrix \footnote{Since the coupling matrix is anti-symmetric,
	it eigenvalues occur in pairs as $\pm \elambda_k$. This affects the level correlations
	close to $\elambda=0$, but they converge rapidly to the level correlations of the Ginibre
	ensemble away from $\elambda=0$. These correlations do not enter in the free energy discussed
	below.}.
To be concrete, the Hamiltonian $H_L$ in \eqref{eq:sykq2} becomes
\begin{equation}
H_L = \sum_{k=1}^{N/4} \elambda_k (2\tilde c_k c_k -1),
\label{h-ff}
\end{equation} 
where $\elambda_k$ are the eigenvalues of  the coupling matrix
$i(J_{ij}+ i k M_{ij})/2$ with positive real parts and hence the sum runs up to $k=N/4$ instead of $N/2$.
Just as in the Hermitian case \cite{Cotler:2016fpe},  we have 
\begin{equation}
[H_L, \tilde c_k] =\elambda_k \tilde c_k, \quad [H_L,  c_k] =-\elambda_k c_k.
\end{equation}
with
\begin{equation}
\tilde{c}_k^2=0, \quad,  c_k^2 = 0,  \quad \{c_k,\tilde c_l\} = \delta_{kl}.
\end{equation}
Hence, we conclude the many-body energies of $H_L$ are given by  filling  $N/4$ free fermions into the single particle states of \eref{h-ff}.
We note that generally $c_k$ is not the Hermitian conjugate of $\tilde c_k$, and hence the energy eigenstates are not necessarily orthogonal to each other,  which is consistent with the nonhermiticity of the Hamiltonian.

In this free-fermion representation, the quenched free energy of the
one-site SYK model, which we shall see to be identical to half the two-site model, is simply
 \begin{equation}\label{eqn:q2Quenched}
 \begin{split}
 F_L=-T\langle \log Z_L\rangle &=-T\left \langle \sum_{k=1}^{N/4}  \log \left(e^{- \elambda_k /T}+ e^{ \elambda_k /T}\right)\right\rangle \\
 & =-\frac{T}{2} \int d^2 z \langle \rho(z )\rangle   \log \left(2\cosh \frac{z}{T}\right),
 \end{split}
 \end{equation}
 where  $\langle \rho(z)\rangle$ is the averaged spectral density of the coupling matrix $i(J_{ij}+ i k M_{ij})/2$.     Notice that since $\elambda_k$ are only half of the levels of $i(J_{ij}+ i k M_{ij})/2$ (those with positive real parts), the integral in the second equality should have only covered half of the support of $\rho(z)$. However, since the integrand is invariant under $z\mapsto -z$, we simply integrate over the whole support and compensate it by a pre-factor of $1/2$.

 At large $N$, the averaged spectral density $\langle  \rho(z )\rangle$ is a constant inside the ellipse \cite{fyodorov1997, hastings2001, hamazaki2020, akemann2022spacing} 
 \begin{equation}\label{eqn:specDenEllipse}
 \rho(z) = \frac{N/2}{\pi \E_0 \y_0} \theta\left(1- \frac{x^2}{\E_0^2} -\frac{y^2}{\y_0^2} \right),
 \end{equation}
 where
 \begin{equation}\label{eqn:realAndImEn}
   \epsilon_0 = \frac {1}{\sqrt{1+k^2}}v, \quad
   \upsilon_0 =\frac{k^2 }{\sqrt{1+k^2}}v
 \end{equation}
and $v$ is the physical scale introduced in equation \eqref{eqn:variancePhysical}.

For $k=0$ the eigenvalues are real with spectral density given by
 \be
 \rho(x) = \frac N{\pi \epsilon_0^2} \sqrt{\epsilon_0^2 -x^2}.
 \ee
 The free energy was already calculated before \cite{maldacena2016} and is
 given by
 \be
-\frac{\beta F}{N/2}=   \frac T{\pi\epsilon_0^2} \int_{-\epsilon_0}^{\epsilon_0}  \sqrt{\epsilon_0^2 -x^2}
 \log \left(2\cosh \frac{x}{T}\right).
 \ee
 Using the Weierstrass product formula for $\cosh x$, this can be expressed
 as
 \be
-\frac{\beta F}{N/2}= \frac 1 {\pi \epsilon_0^2} \int_{-\epsilon_0}^{\epsilon_0}  \sqrt{\epsilon_0^2 -x^2}
\left ( \log 2 +  \sum_{n=0}^\infty \log \left( 1+
\frac {4 x^2}{\omega_n^2}\right)\right ),
 \ee
 where $\omega_n$ are the Matsubara frequencies
 \be
 \omega_n = \frac{2\pi(n+\frac 12 )}{\beta}.
\label{mats}
 \ee
 The integral over $x$ is known analytically resulting in
\be\label{eqn:hermitianFreeEn}
 -\frac{\beta F}{N/2}= \frac{1}{2}\log 2+\sum_{n\geq 0}\left\{\log \left [ \frac 12
   + \frac 12 \sqrt{1+ \frac {4 v^2}{\omega_n^2}}\right ]+
 \frac{\sqrt{1+4 v^2/\omega_n^2}-1 -4v^2/\omega_n^2}{4 v^2/\omega_n^2}\right\}.
   \ee
   We recognize the expressions for the Green's function and the self-energy in the $\Sigma G $ formulation \cite{Cotler:2016fpe} suggesting that this result can also be obtained from the
   solution of the SD equations, see section \ref{sec:sd}.

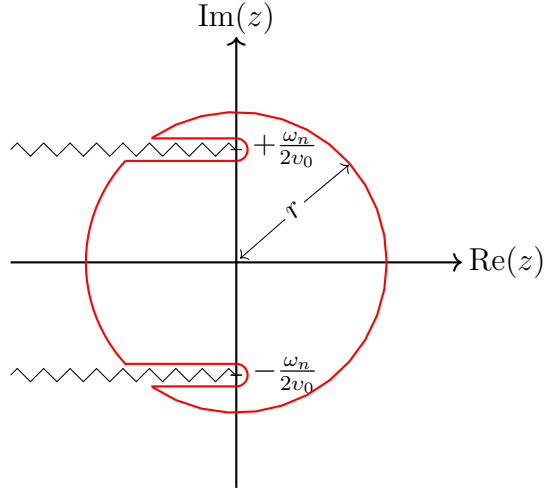
\begin{figure}
	\begin{center}
		\begin{tikzpicture}[scale =0.5]
		\draw[thick,->] (-6,0) -- (6,0);
		\draw[thick,->] (0,-6) -- (0,6);
		\node at  (7.2,0) {Re$(z)$};
		\node at  (0,6.5) {Im$(z)$};
		
		\node[anchor=west] at (0,3) {$\,+\frac{\omega_n}{2\y_0}$};
		\draw (-0.15, 3) -- (0.15, 3);
		\draw[snake=zigzag]     (-6,3) -- (0,3);
		
		\node[anchor=west] at (0,-3) {$\,-\frac{\omega_n}{2\y_0}$};
		\draw (-0.15, -3) -- (0.15, -3);
		\draw[snake=zigzag]     (-6,-3) -- (0,-3);
		
		\draw (-0.15, -3) -- (0.15, -3);
		\draw [->] (1.8,1.6) -- (3,2.6);
		\draw [->] (1.3,1.15) -- (0.1,0.1);
		\node at (1.55, 1.4) {\rotatebox{45}{$r$}};
		
		\draw [red,thick,domain=-124:124] plot ({4*cos(\x)}, {4*sin(\x)});
		\draw [red, thick] (-2.26,3.3) --(0,3.3);
		\draw [red, thick] (-2.26,-3.3) --(0,-3.3);
		
		\draw [red, thick] (-2.95,2.7) --(0,2.7);
		\draw [red, thick] (-2.95,-2.7) --(0,-2.7);

		\draw [red,thick,domain=-90:90] plot ({0.3*cos(\x)}, {0.3*sin(\x)+3});
		\draw [red,thick,domain=-90:90] plot ({0.3*cos(\x)}, {0.3*sin(\x)-3});

		\draw [red,thick,domain=137.5:223] plot ({4*cos(\x)}, {4*sin(\x)});

		\end{tikzpicture}
	\end{center} \caption{A contour that gives $I_n(r) +$branch cut contributions. The integrand of \eqref{eqn:contourMatsubara} is analytic in the interior of this contour.}
        \label{fig:contour}
\end{figure}

 Next we  discuss the case $k = 1$ where the ellipse becomes a circle
 with $\epsilon_0 = \upsilon_0$ (for the general elliptic case see appendix \ref{a:yiyangfreefermions}).  The  free energy \eqref{eqn:q2Quenched} can be expressed as
\begin{equation}\label{eqn:freeEnIntegral}
\begin{split}
-\frac{F_L}{(N/4)T} &= \frac{1}{\pi \upsilon_0^2}\int_{D_{\upsilon_0}} d^2 z   \log \left(2\cosh  \frac{z}{T}\right)\\
&  = \int_0^1 r  I(r) dr ,
\end{split}
\end{equation}
where $D_{\upsilon_0}$ represents a disk of radius $\upsilon_0$ centered at the origin, and in the second equality we have scaled the integral to the unit disk where
\begin{equation}\label{eqn:contourInt}
\begin{split}
I(r) = &\frac{1}{\pi } \int_0^{2\pi} d \phi \log \left( 2\cosh \frac{\upsilon_0 r \cos \phi + i \upsilon_0r \sin \phi}{T}\right)\\
=&\frac{1}{\pi i}\oint_{S_r} \frac{dz}{ z}  \log \left(2\cosh \frac{\upsilon_0 z}{T}\right),
\end{split}
\end{equation}
where  $S_r$ is a circle of radius $r$.  We can directly evaluate $I(r)$ by
expressing the cosh as a product over Matsubara frequencies using the Weierstrass
formula,
\begin{equation} \label{eqn:circularInfSum}
I(r)=2\log 2+ \sum_{n=0}^\infty I_n(r),
\end{equation}
where 
\begin{equation}\label{eqn:contourMatsubara}
I_n(r) = \frac{1}{\pi i} \oint_{S_r} \frac{dz}{z}  \log \left(1+ \frac{4\y_0^2 z^2}{\omega_n^2}\right).
\end{equation}
We note that the integrand of $I_n$ has two cuts and no pole: the two cuts start at $\pm i  \omega_n/2\y_0$ and extend horizontally to the negative infinity.  If the circle $S_r$ does not touch the cuts ($r<\omega_n/2\y_0$),  then $I_n =0$; if $S_r$ intersects with the cuts ($r>\omega_n/2\y_0$),  then we choose the contour to narrowly avoid the cuts (see figure \ref{fig:contour}),
then we have that the sum of $I_n$  and the cut contributions vanishes,
so that the problem reduces to evaluating the contributions from the parts that surround the cuts.  Hence we conclude 
\begin{equation}
  I_n(r) =\begin{cases}
\displaystyle
-2  \int_{-\sqrt{r^2 - \omega_n^2/4\y_0^2}}^0 dx\left(\frac{ 1 }{x + i \omega_n/2\y_0}+\frac{ 1 }{x - i \omega_n/2\y_0}\right)= 2 \log \left(\frac{4\y_0^2 r^2}{\omega_n^2}\right) \quad \text{if $\omega_n < 2\y_0 r$},\\
\displaystyle
0 \hspace{9.75cm}\text{if $\omega_n > 2\y_0r.$}
\end{cases}
\end{equation}
Equation \eqref{eqn:circularInfSum} finally truncates to
\begin{equation}
I(r) =2\log 2 +2 \sum_{0<\omega_n<2\y_0 r}\log\left(\frac{4\y_0^2 r^2}{\omega_n^2}\right).
\end{equation}
The free energy in \eqref{eqn:freeEnIntegral} is then given by
\begin{equation}
\begin{split}
-\frac{F_L}{(N/4)T}=& \int_0^1 r I(r) dr\\ 
=& \log 2 + \int_{\frac{\omega_n}{2\y_0}}^1 \sum_{0<\omega_n<2\y_0 r}  2r  \log\left(\frac{4\y_0^2r^2}{\omega_n^2}\right)dr\\ 
=&\log 2 + \sum_{0<\omega_n<2\y_0 } \left[-1 +\frac{\omega_n^2}{4\y_0^2}-  \log \frac{\omega_n^2}{4\y_0^2} \right]. \\
\end{split}
\label{free-ev}
\end{equation}
In section \ref{sec:sd} we will show that this result can also be obtained
from the solutions of the Schwinger-Dyson equations.

 \begin{figure}[t!]
 	\centerline{
 		\includegraphics[width=8cm]{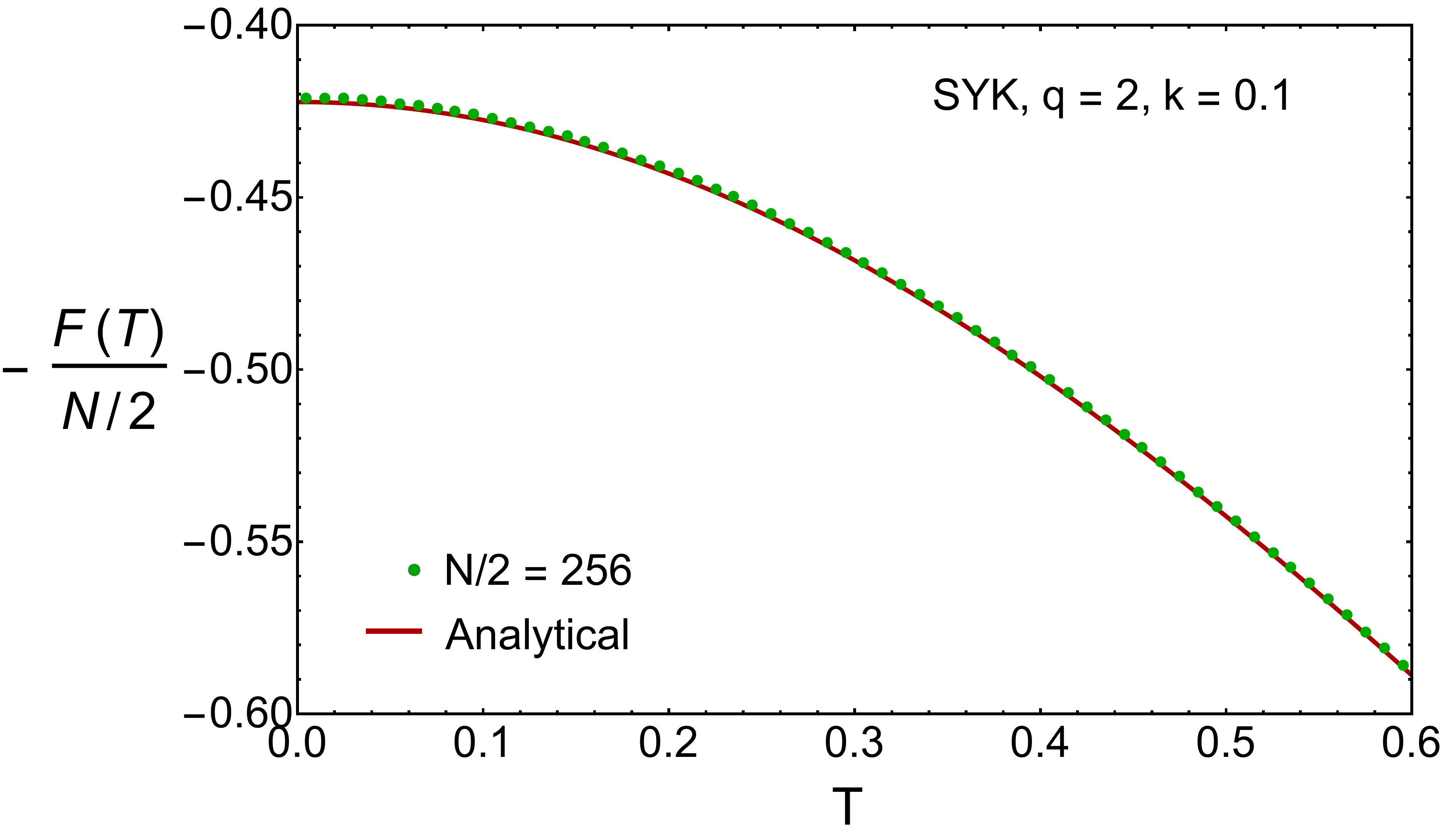}
		\includegraphics[width=8cm]{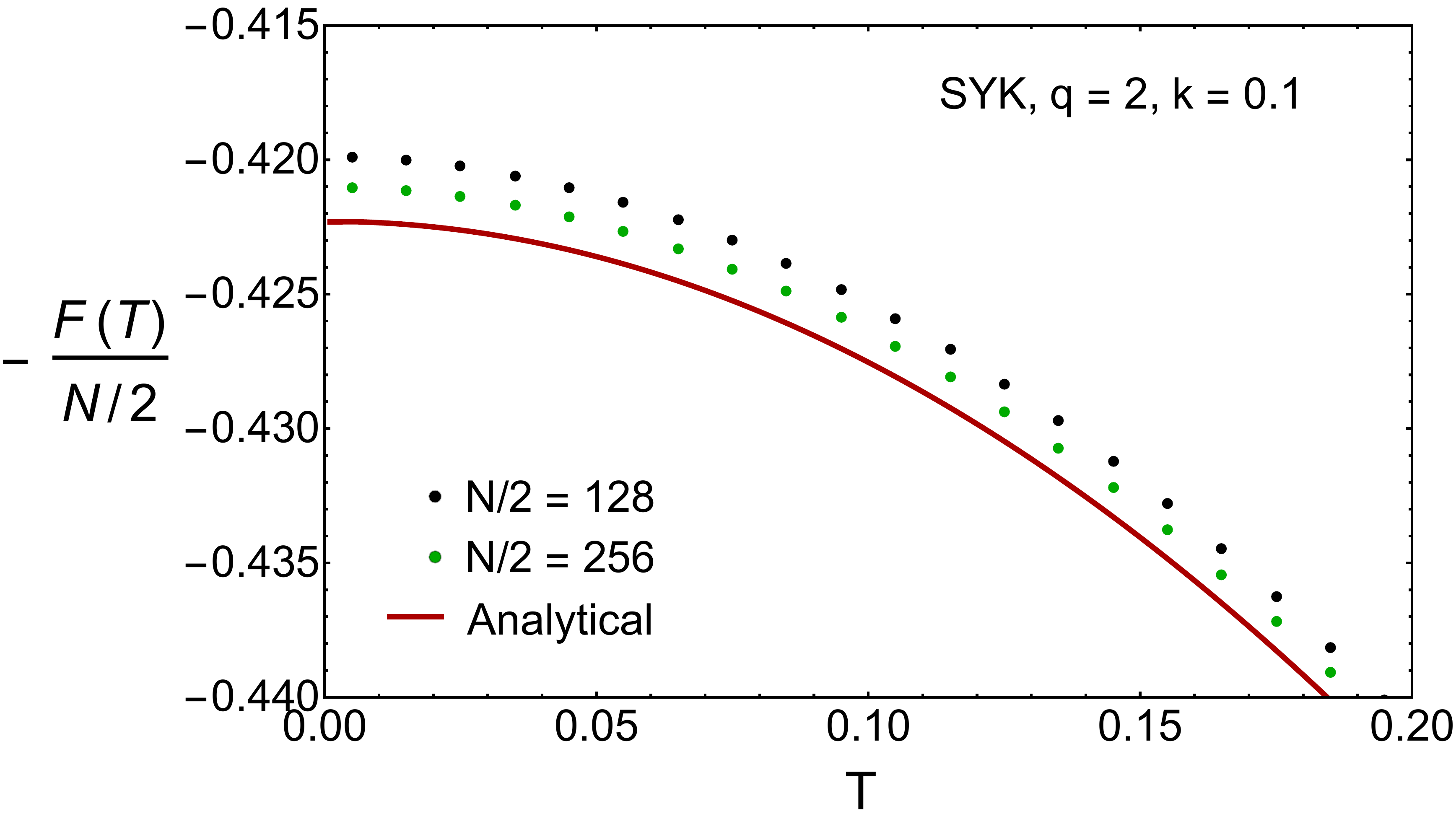}
 	}
 	\centerline{
 		\includegraphics[width=8cm]{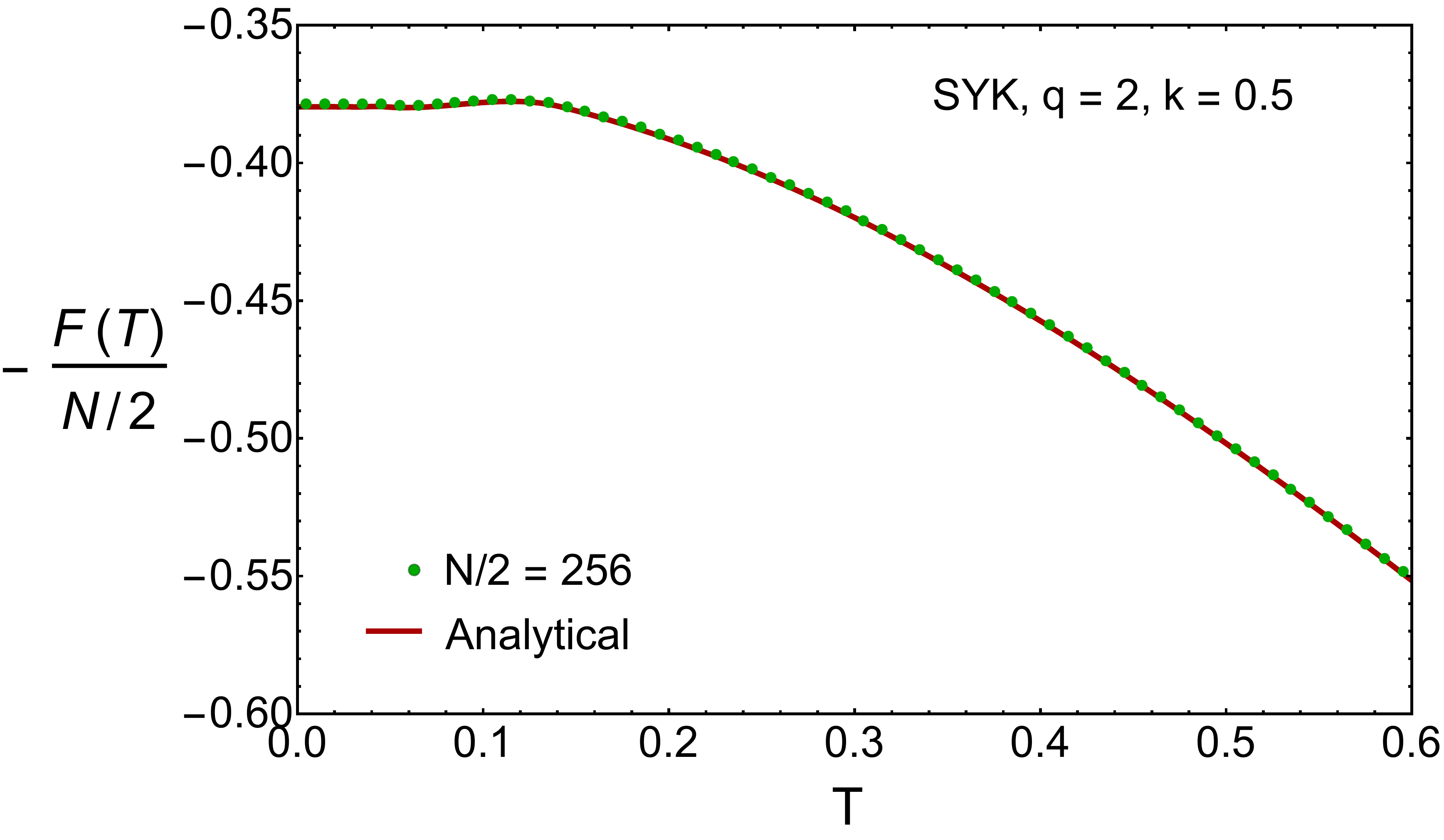}
		\includegraphics[width=8cm]{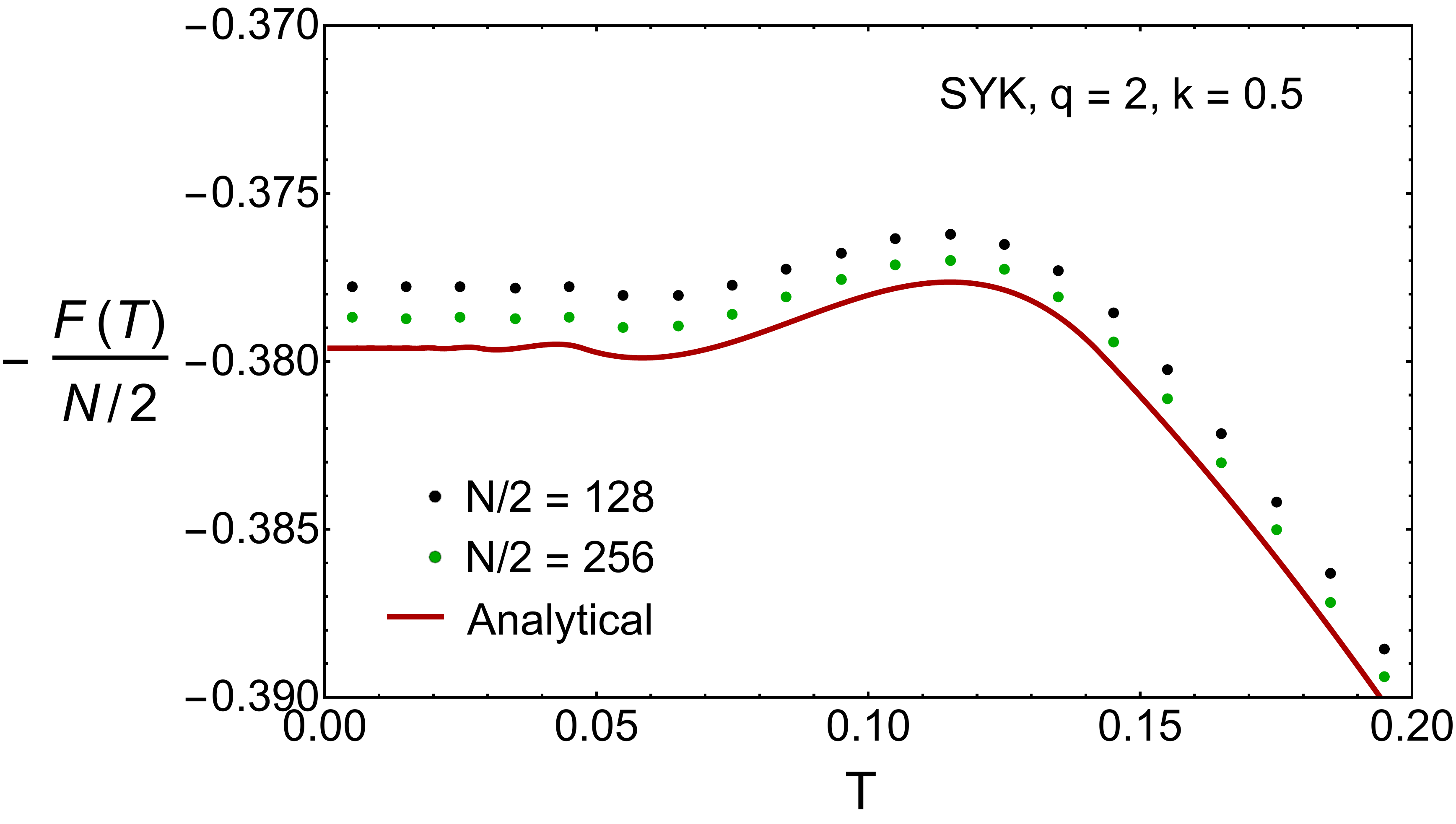}
 	}
 	\centerline{
 		\includegraphics[width=8cm]{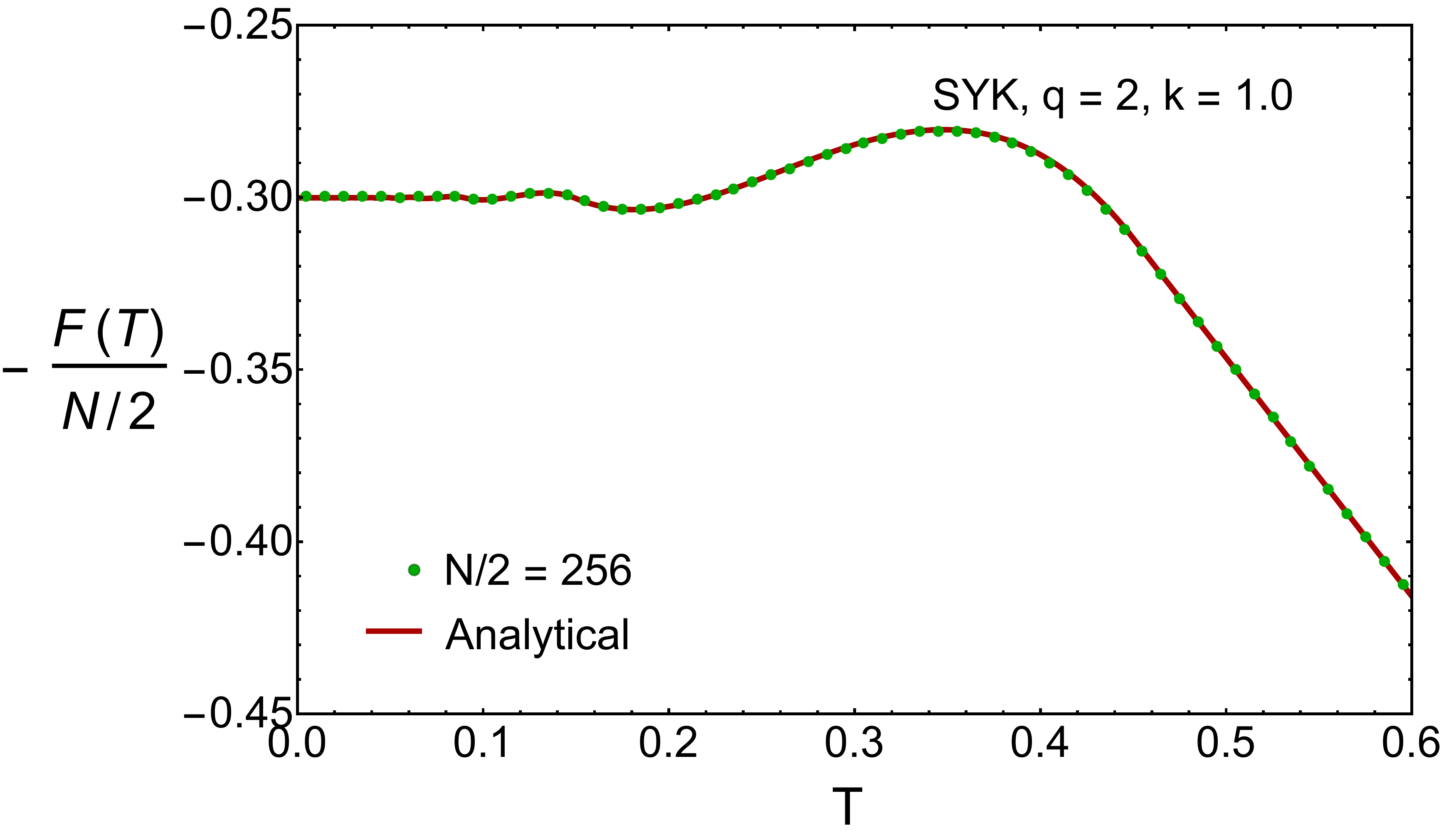}
		\includegraphics[width=8cm]{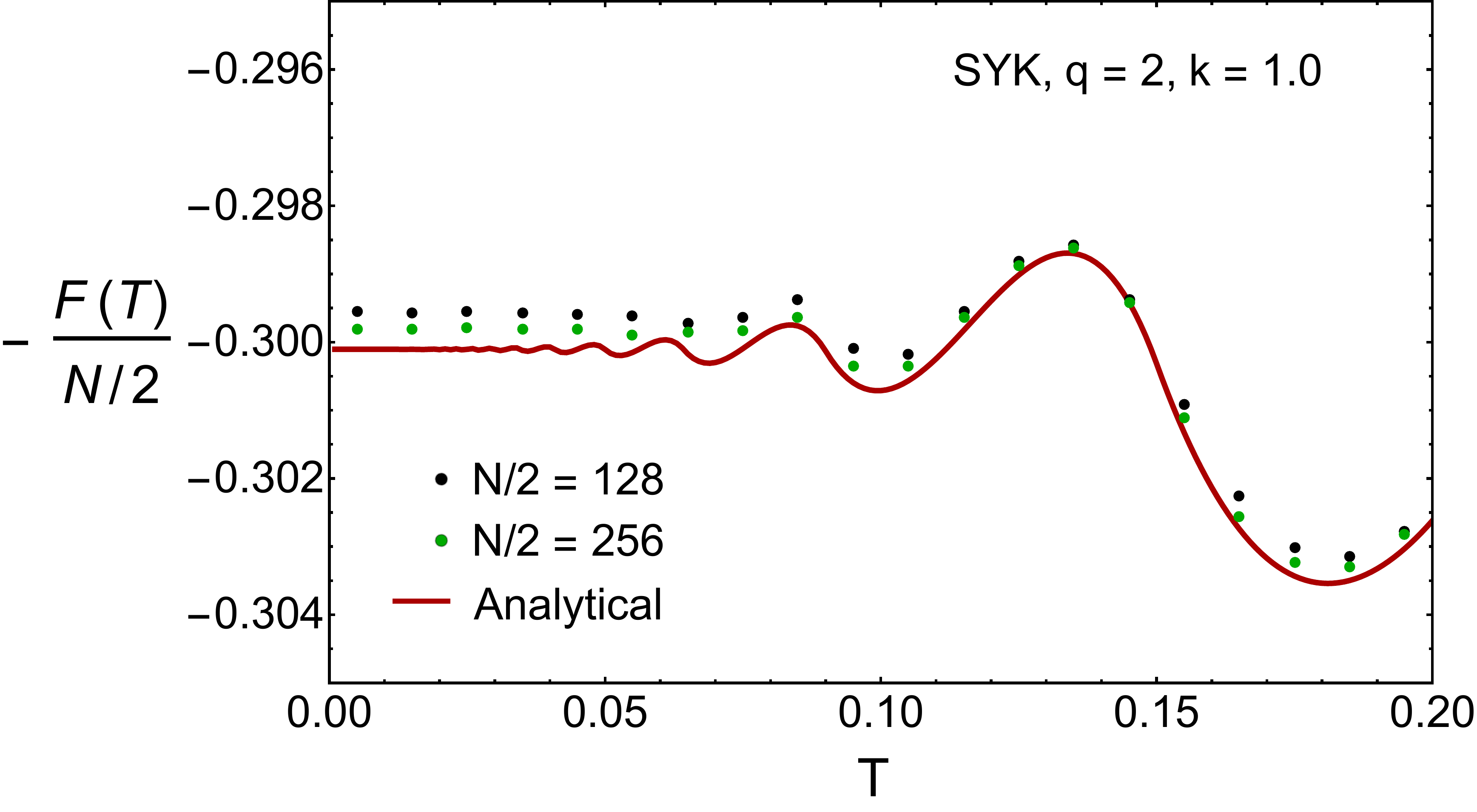}
 	}

 	\caption{The free energy of the $q=2$ SYK model for $k=0.1$ (top),
          $k=0.5$ (middle) and $k=1$ (bottom).
          The dots
  denote the results obtained by averaging over an ensemble of 1000
  configurations for $N/2=128$ (black) and $N/2=256$ (green) Majorana fermions.
  The right figure is magnification of the left figure. The $N/2=128$ data are
  only shown in the right figure. The red curve represents the analytical
  result for the free energy.
}
 	\label{fig:free-q2}
 \end{figure}

 We observe that a phase transition occurs at temperatures for which an additional Matsubara frequency
 enters in the sum of equation \eref{free-ev}. This happens for $\frac{\y_0}{\pi T} -\frac{1}{2} =m $ ($m$ being non-negative integers) resulting
 in a series of  
 critical temperatures parameterized by $m$:
 \be\label{tcm}
 T_{c,m} = \frac {2\y_0/\pi}{2m+1},  
 \ee
 where $\y_0 = v/\sqrt{2}$ according to equation \eqref{eqn:realAndImEn}. For $T> T_{c,0}$, there are no Matsubara frequency satisfying $0<\omega_n<2\y_0$,
 and no more phase transitions can occur.

 We re-iterate that, because of the complex-conjugation invariance of the
 {\it average} single-particle spectral density,  we have 
 \begin{equation}
 \langle \log Z_L\rangle =\langle \log Z_L^*\rangle ,
 \end{equation}
 so that the two-site quenched free energy just doubles that of the one-site and the free energy density remains the same, that is,  $F/N = F_L/(N/2)$.
 
\begin{figure}[t!]
 	\centerline{
 		\includegraphics[width=8cm]{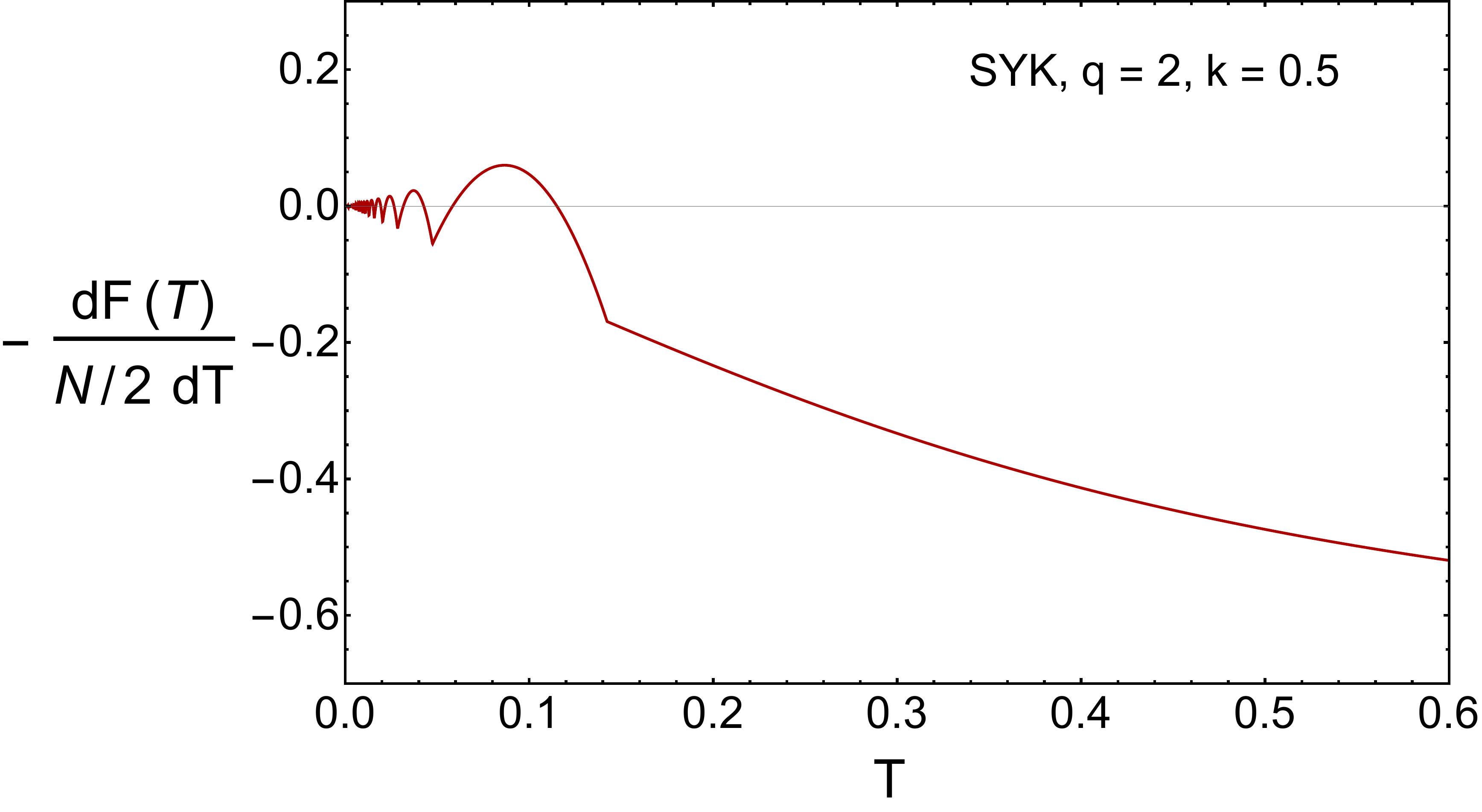}
 		\includegraphics[width=8cm]{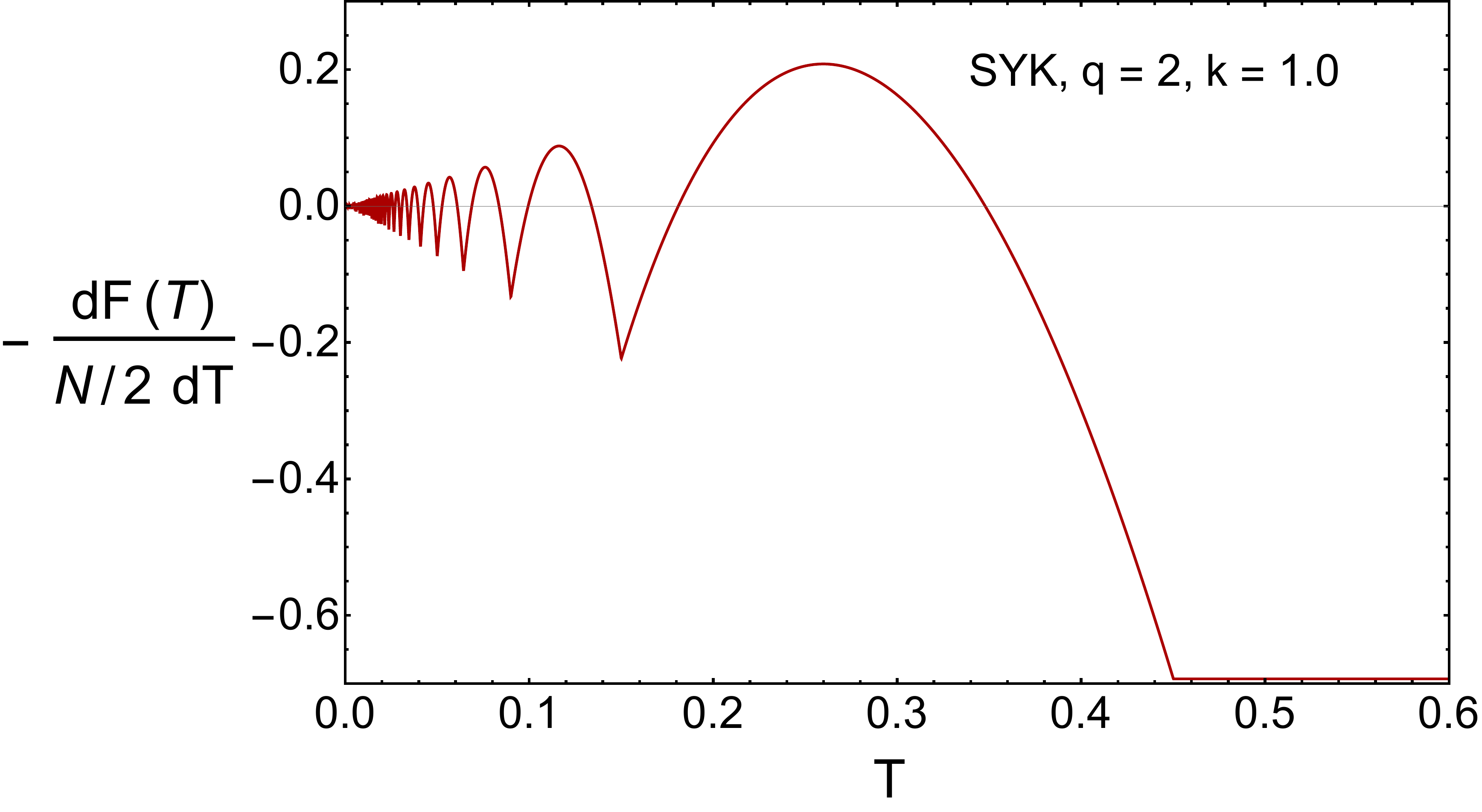}}
 	\caption{Derivative of the free energy per particle of the non-Hermitian $q=2$ SYK model for  $k=0.5$ (left) and $k=1$ (right). The kinks, which indicate the existence of second-order phase transitions, are located at the temperatures (\ref{tcm}) where a new Matsubara frequency contributes to the free energy (\ref{free-ff}).
          The last kink appears at a temperature $\sim k^2$ and would be barely visible
          for $k=0.1$ which is not shown.
          The physical scale is set to $v=1$.}
 	\label{fig:dfree-q2}
 \end{figure}

 To determine the order of the phase transition 
it is useful to  study the first derivative of the free-energy.
Indeed, we observe, see figure ~\ref{fig:dfree-q2}, that it has  kinks that point toward a family of second-order phase transitions,  
 each time
 a new pair of Matsubara frequencies enters in the sum. This can be shown
 analytically by
 expanding the free energy around $T_{c,m}$. We find that the  contribution at this new critical temperature scales as $\sim(T-T_{c,m})^2$ so, as depicted in figure \ref{fig:free-q2}, the free energy \eref{free-ev} is smooth around this critical temperature and therefore the transition cannot be of first order. However, one can easily show that the derivative of the free energy is not smooth
 at $T_{c,m}$ as is also clear from figure
 \ref{fig:dfree-q2}.
 
 Using similar methods, the integrals can also be calculated for $0<k<1$, see
 appendix \ref{app:b},
 \be
-\frac{F_L}{(N/2)T}    
&=&\frac 12\log 2  + \sum_{\omega_n \geq 2\y_0}^\infty
\left\{ \log \left[\frac{1}{2}\left(1 + \sqrt{1+\frac{\E_0^2-\y_0^2}{\omega_n^2/4}}\right) \right] -\frac 12 \frac{\omega_n^2/4}{\E_0^2-\y_0^2}\left(1-\sqrt{1+\frac{\E_0^2-\y_0^2}{\omega_n^2/4}}\right)^2     \right\}\nn \\
&&  + \frac 12 \sum_{0<\omega_n<2 \y_0}\left\{ \log \frac{(\E_0+\y_0)^2}{\omega_n^2}  - 1 - \frac{\omega_n^2(1-\E_0/\y_0)}{2(\E_0^2 - \y_0^2)}\right\}.
\label{free-ff}
\ee
Each term  contributing to the sum in the first line is equal 
to the corresponding term in  the free energy of the Hermitian SYK model \eqref{eqn:hermitianFreeEn} with $
\epsilon_0 \to \sqrt{\epsilon_0^2 -\y_0^2}$. In figures \ref{fig:free-q2}
and \ref{fig:dfree-q2} we show the temperature dependence of the free energy
and its derivative for various values of $k$. It is disturbing that the
entropy becomes negative which cannot be due to the failure of the replica
trick since it is not used in free-fermion method.  Most likely it is a consequence of the nonhermiticity which will become
more clear in section \ref{sect:sigma}.

The critical temperatures are given by the same expression as for $k=1$,
\begin{align}
 \label{eq:critical_temperature_k_less_1}
 T_{\mathrm{crit}}^{(n)} = \frac{2v_0/\pi}{ 2n + 1 } \ ,\qquad, n = 0,1,2, \cdots 
 \end{align}
but with $\y_0 = k^2v/\sqrt{1+k^2}$ according to equation \eqref{eqn:realAndImEn}.   One can easily show that the derivative of the free energy is continuous
 at the critical points while its second derivative is discontinuous.
 The appearance of an infinite number of critical points is a direct
 consequence of the factorization of the partition function into a
 product over positive Matsubara frequencies.
 For each Matsubara frequency we have
 exactly one critical temperature.

    \section{Free energy from the Schwinger-Dyson equations}
 \label{sec:sd}
 
 We now turn to confirm these results by an explicit large $N$ calculation from the solutions of the Schwinger-Dyson (SD) equations of the SYK model. In this case,
 the calculation is also analytical, though very different from the one carried out in the previous section. However, we shall see that ultimately the expression for the free energy is the same.  In the Schwinger-Dyson approach,
  replica symmetry breaking between conjugate replicas
  plays a crucial role. 
  Indeed, we shall see that for $k=1$ the free energy is determined by the Green's function $G_{LR}$ (equivalently its self energy $\Sigma_{LR}$) related
  to the effective coupling of the two sites. However, it is assumed that the replica symmetry of a conjugate pair
  remains unbroken so that the quenched free energy can be obtained from
  just one replica and one conjugate replica.
 
The Euclidean $\Sigma G$ action for the $q$-body SYK model takes the form \cite{maldacena2018}
\be
	\label{eq:MQ_model_action}
        - \frac{2 S_E}{N} &=& \log \mathrm{Pf}(\partial_t \delta_{ab} - \Sigma_{ab}) - \frac 12  \int d\tau_1 d\tau_2 \sum_{a , b} \left[\Sigma_{ab}(\tau_1 , \tau_2) G_{ab}(\tau_1 , \tau_2) - s_{ab} \frac{\mathcal J_{ab}^2}{2 q^2} [2 G_{ab}(\tau_1 , \tau_2)]^q \right]
        \nn\\
        &&-\frac i2\epsilon \int d\tau(G_{LR}(\tau,\tau) -G_{RL}(\tau,\tau))
        \ ,
\ee
where the indices $a , \, b$ can be equal to $R$ or $L$.  The integrations over $\tau$ variables are on the interval $[0,\beta]$. The factor $s_{ab}$ is equal
to 1 for $a=b$ and equal to $(- 1)^{q/2}=-1$ for $a\ne b$. 
The couplings take the value  ${\mathcal J}_{LL}={\mathcal J}_{RR}={\mathcal J}$ when $a = b$ and ${\mathcal J}_{LR}={\mathcal J}_{RL}=\mathcal{\widetilde J}$ when $a \neq b$. The term proportional to $\epsilon$ is included to break the symmetry
that requires $G_{LR}$ to vanish
(more details are given in the analysis of the $q=4$ model\cite{Garcia-Garcia:2022}).
In terms of   the random coupling couplings of the $L$ and $R$ gamma matrices,
$J_{ij}^L$ and $J_{ij}^R$,
for $q=2$ the constants  ${\mathcal J}$ and $\widetilde {\mathcal J}$ are defined by
\begin{align}
\label{eq:variances_mine}
&\mathcal J^2 : =\frac{N}{2}  \langle (J^L_{ij})^2 \rangle = \frac{N}{2}\langle (J^R_{ij})^2 \rangle =  (1-k^2)v^2 \ , \nn \\
&\widetilde{\mathcal J}^2 :=  \frac{N}{2}\langle J^L_{ij}   J^R_{ij}\rangle =   (1+k^2)v^2 \ ,
\end{align}
where the left and right couplings are related to the couplings $J_{ij}$ and $M_{ij}$ by
\be
\label{eq:coupling_mine_antonio_mapping_definition}
& J^L_{ij} \equiv J_{ij} +  i \, k M_{ij} \ , \nn \\
&J^R_{ij} \equiv J_{ij} -  i \, k M_{ij} \ .
\ee
For a discussion of the symmetries of $G_{ab}\sim \Sigma_{ab}$
we refer to a study of the $q=4$ model \cite{Garcia-Garcia:2022}.
As it stands the integrals over $\Sigma$ and $G$ for the action
    \eref{eq:MQ_model_action}
    are not convergent which is required to perform the integrations over
    $G_{ab}$.
    Convergence can be achieved by rotating
    \be
    G_{RR}(\tau_1,\tau_2) \to i G_{RR}(\tau_1,\tau_2),\qquad
    G_{LL}(\tau_1,\tau_2) \to i G_{LL}(\tau_1,\tau_2)\nn\\
    \Sigma_{RL}(\tau_1,\tau_2) \to i \Sigma_{RL}(\tau_1,\tau_2), \qquad
    \Sigma_{LR}(\tau_1,\tau_2) \to i \Sigma_{LR}(\tau_1,\tau_2).
    \ee
    The rotations do not affect the saddle-point evaluation of the action integral, but as we shall see below, explicitly integrating out $G$ for $q=2$ simplifies the saddle-point analysis and we prefer to use a convergent definition for this reason.   The action that gives a convergent path integral is then
 \be
	\label{eq:MQ_model_actionRotated}
        - \frac{2 S_E}{N} &=& \log \mathrm{Pf}(\partial_\tau \delta_{ab} - \xi_{ab}\Sigma_{ab}) - \frac 12  \int d\tau_1 d\tau_2 \sum_{a b} \left[i\Sigma_{ab}(\tau_1 , \tau_2) G_{ab}(\tau_1 , \tau_2)+ \frac{\mathcal J_{ab}^2}{2 }  G_{ab}(\tau_1 , \tau_2)^2 \right]
        \nn\\
        &&-\frac i 2\epsilon \int d\tau(G_{LR}(\tau,\tau) -G_{RL}(\tau,\tau)) 
        \ ,
        \ee
        with $\xi_{LL}=\xi_{RR} =1 $ and
        $\xi_{LR}=\xi_{RL} = i $.
Contrary to  $q=4$, the integrals over $G_{ab}(\tau_1,\tau_2)$ are Gaussian and can be carried 
out exactly. This
    results in the effective action for $\Sigma_{ab}$ 
\be
	\label{eq:MQ_model_actionSigmaOnly}
        - \frac{2 S_E(\Sigma)}{N} &=& \log \mathrm{Pf}(\partial_\tau \delta_{ab} - \xi_{ab}\Sigma_{ab}) - \frac 14  \int d\tau_1 d\tau_2 \sum_{a , b}
        \left(\frac{\Sigma_{ab}(\tau_1 , \tau_2)}{\mj_{ab}} \right)^2\nn\\ 
        && -\frac 12\epsilon \int d\tau(\Sigma_{LR}(\tau,\tau) -\Sigma_{RL}(\tau,\tau)), 
        \ .
        \ee

        At the saddle point, $\Sigma$ should be translation-invariant and hence only a function of $\tau_1-\tau_2$.  For the purpose of saddle-point analysis, we can express the action in terms of Fourier modes of the $\Sigma_{ab}$ which is now a single-variable anti-periodic function:
\be
 \label{eq:fourier_expansion_GLR_k1}
 \Sigma_{ab}(\tau) &=& \frac 1\beta \sum_{ \omega_n}
e^{-i\omega_n \tau} \Sigma_{ab}(\omega_n),
\ee
where $\omega_n =  (2n+1)\pi/\beta$ are Matsubara frequencies
 defined already in equation \eqref{mats}.
 This results in
\be
	\label{S-sig-om}
        - \frac{2 S_E(\Sigma)}{N} &=& \sum_{\omega_n}
        \frac12 \log \det(-i\omega_n \delta_{ab}-  \xi_{ab}\Sigma_{ab}(\omega_n)) - \frac 14  \sum_{a b}
        \frac{\Sigma_{ab}(\omega_n)\Sigma_{ab}(-\omega_n)}{\mj_{ab}^2 } \nn\\
          &&-\frac 12\epsilon \sum_{\omega_n}(\Sigma_{LR}(\omega_n)
          -\Sigma_{RL}(\omega_n))
        \ .
\ee
 As already emphasized, contrary to the $q=4$ case, in the $q=2$ case the SD equation simplify to second order equations (we took the limit $\epsilon \to 0$)
 \be
   \label{eq:SD_equations_quadratic}
   \frac{i\omega_n+\Sigma_{RR}(\omega_n)}
        {(i\omega_n+\Sigma_{LL}(\omega_n))(i\omega_n+\Sigma_{RR}(\omega_n))
             +\Sigma_{LR}(\omega_n) \Sigma_{RL}(\omega_n)}&=&
   \frac{\Sigma_{LL}(-\omega_n)}{\mj^2},\nn\\
   \frac{\Sigma_{RL}(\omega_n)}
         {(i\omega_n+\Sigma_{LL}(\omega_n))(i\omega_n+\Sigma_{RR}(\omega_n))
             +\Sigma_{LR}(\omega_n) \Sigma_{RL}(\omega_n)}&=&
   \frac{\Sigma_{LR}(-\omega_n)}{\mjt^2},
     \ee
and anther two equations with subscripts $L$ and $R$ interchanged. At the saddle point we have that $\Sigma_{RL}(\omega_n) =-\Sigma_{LR}(\omega_n)$ and  
$\Sigma_{RR}(\omega_n) =\Sigma_{LL}(\omega_n)$. The saddle point equations
couple positive and negative frequencies, but the solutions
 are simply related by
\be
\Sigma_{LL}(-\omega_n) = - \Sigma_{LL}(\omega_n), \qquad
\Sigma_{LR}(-\omega_n) =  \Sigma_{LR}(\omega_n).
\ee
Using these relations the saddle point equations are easily solved with
a trivial solution  given by (the symmetries of $G_{ab}$ and  $\Sigma_{ab}$
are discussed in detail in the analysis of  the $q=4$ model \cite{Garcia-Garcia:2022}),
\be
\label{sad-t}
\Sigma_{LR}(\omega_n) &=& 0,\nn\\
\Sigma_{LL}(\omega_n) &=& -\frac i2\omega_n \pm \frac i2{\sign}(\omega_n)\sqrt{4\mj^2 + \omega_n^2},
\ee
and a nontrivial solution that couples the Left and Right SYK models breaking
the replica symmetry between them:
\be
\label{sad-nt}
\Sigma_{LR}(\omega_n)&=&\pm \mjt \sqrt{1- \frac{\omega_n^2}{{\omega_{\rm cr}}^2}},\nn\\
\Sigma_{LL}(\omega_n)&=& \frac{ i\mj^2\omega_n}{\mjt^2-\mj^2},
\ee
where we have introduced the critical frequency
\be\label{om-cr}
\omega_{\rm cr} = \frac{\mjt^2-\mj^2}\mjt\; .
\ee
  This frequency will play an important role in the analysis of the partition
  function.

  Pairing the positive and negative Matsubara frequencies allows us to write the free energy as a sum over only positive  Matsubara frequencies:
\begin{equation}
-\frac{2F}{NT}= \sum_{\omega_n>0}\left\{\frac{1}{2}\log \left[\left( i \omega_n+\Sigma_{LL}(\omega_n)\right)^2 -\Sigma_{LR}(\omega_n)^2\right]^2+\frac{\Sigma_{LL}(\omega_n)^2}{\mj^2}-\frac{\Sigma_{LR}(\omega_n)^2}{\mjt^2}\right\}.
\end{equation}
For unbroken saddles we have two solutions, and it turns out  one solution gives a larger $-S_E$ hence is the dominant saddle. This is the solution with $+\sign(\omega_n)$ term for $\Sigma_{LL}$.    The free energy
  of this solution is equal to the free energy
  of two uncoupled SYK models \cite{Cotler:2016fpe} (but with $\mj^2 =(1-k^2)v^2$).     For each (positive) Matsubara frequency this gives
\be
-\frac{2F_{\rm \;2\;SYK}(\omega_n)}{NT} =  -\frac{\omega_n^2}{4\mj^2} \left(1-
 \sqrt{1 + \frac{4\mj^2}{\omega_n^2}}
\right)^2+ \log\left[ \frac 14 \left(1 + \sqrt{ 1+ \frac{4\mj^2}{\omega_n^2}}\right)^2\right]
    + \log (\omega_n^2),\nn\\
    \ee
    while the free energy of the broken solution reduces to
\be
-\frac{2F_{\rm Broken}(\omega_n)}{NT} = -1 + \frac {\omega_n^2}{\mjt^2-\mj^2}
+  \log \mjt^2.
\ee
The broken solution always gives the dominant action, but as we will see next,
it does not always determine the free energy.
The saddle point of $\Sigma_{LL}(\omega_n)$ is always purely imaginary, but the
saddle point of $\Sigma_{LR}(\omega_n)$ switches from real to imaginary at
 $\omega_n=\omega_{\rm cr}$ where the free energy of the trivial and
the nontrivial
solution coincides. For $\omega_n >\omega_{cr}$
the imaginary part
of the action is zero at the saddle point, but the action becomes
complex along the integration manifold.
In order to apply the steepest descent method, the integration manifold must
be directed along the Picard-Lefschetz thimble. Otherwise we will have large
cancellations that may suppress the action of the saddle point
minimizes the  the free energy. It is a complicated problem
to find the Lefschetz thimbles in a multidimensional space, but we can analyze the problem along the trajectory
 where $\Sigma_{LL}(\omega_n)$ and $\Sigma_{RR}(\omega_n)$ are at the
saddle point while for the off-diagonal $\Sigma_{ab}(\omega_n)$ variables
we restrict ourselves to the sub-manifold
$\Sigma_{RL}(\omega_n)= -\Sigma_{LR}(\omega)_n$
and $\Sigma_{RL}(-\omega_n)= -\Sigma_{LR}(\omega_n)$ which intersects with the
saddle-point. Combining positive and negative Matsubara frequencies, 
the action on this sub-manifold is given by
\be\label{eqn:submanifoldAction}
\frac {-2 S_E(\omega_n)}N = - \frac {\mj^2\omega_n^2}{(\mjt^2-\mj^2)^2}
-\frac{\Sigma_{LR}^2(\omega_n)}{\mjt^2}
+  \log\left (\frac{\mjt^4\omega_n^2}{(\mjt^2-\mj^2)^2}
+\Sigma_{LR}^2(\omega_n) \right ), \qquad \omega_n>0.
\nn\\
\ee
This action also arises in the study of a zero-dimensional Gross-Neveu-like model, and its saddle point analysis
\cite{Kanazawa:2014qma,Tanizaki:2015gpl}, which we will apply here.  
The saddle points of this effective action are still given by the $\Sigma_{LR}$ of 
\eref{sad-t} and \eref{sad-nt} of the full action, namely $\Sigma_{LR}=0$ and $\Sigma_{LR}(\omega_n)=\pm \mjt \sqrt{1- {\omega_n^2}/{\omega_{\rm cr}}^2}$.  At all the saddle points
the action is real and the Lefschetz thimble of these saddle points are the
real axis if the saddle point solution for $\Sigma_{LR}$ is real, and
along the imaginary axis when this
saddle point is imaginary. In the the latter case, i.e. for $\omega_n > \omega_{\rm cr}$, the thimble ends are the zeros of the logarithm, and it is not possible to deform the real axis continuously into the thimble.\footnote{To be precise, to get  well-defined thimbles emanating from the zeros of the logarithm,  the small symmetry-breaking term proportional to $\epsilon$ must be included, and each zero will give a separate thimble.  But the basic conclusion remains the same: these two thimbles do not contribute to the path integral because the original contour of integration cannot be deformed into either of them
\cite{Kanazawa:2014qma,Tanizaki:2015gpl}.}  Of course we can
deform the initial integration over the real axis to an integration path in
the complex plane that goes over the saddle point on the imaginary axis. As long
as we do not cross any singularities, by Cauchy's theorem, the value of the
integral along the deformed path will be the same in spite of the fact that the
integrand at the saddle point on the imaginary axis is much larger than that of the
saddle points on the real axis. The phase
of the integrand together with the Jacobian will assure that  the contributions to
the integral combine to the correct result. However, if we integrate only
over the Gaussian fluctuations about the imaginary saddle point, we do $not$
get the correct result. In order words, we cannot apply the saddle-point
approximation to the imaginary saddle points. Instead, the integral can
be evaluated at
 the trivial saddle point which has
its thimble on the real axis. For $\omega_n < \omega_{\rm cr}$ the
$\Sigma_{LR}(\omega_n)$ integral runs over  both real  saddle-points but one of them is suppressed
by the $\epsilon$ term in the action.  

Strictly speaking, the $\Sigma_{LR}=0$ saddle of the effective action \eqref{eqn:submanifoldAction} does not quite correspond to the $\Sigma_{LR}=0$  saddle of the full action,  because in writing down the effective action we already assumed a $\Sigma_{LL}$ of the form in solution \eqref{sad-nt},  but the $\Sigma_{LR}=0$  saddle of the full action belongs to solution  \eqref{sad-t} where $\Sigma_{LL}$ takes a different form.    Thus $\Sigma_{LR}=0$ solution for the effective action should be viewed as a spurious saddle due to the sub-manifold constraint.
Therefore a more definitive analysis should be performed on the full action (the full action is quite similar, though not exactly the same,  as the action of a  zero-dimensional Nambu-Jona-Lasinio-like model \cite{Kanazawa:2014qma,Tanizaki:2015gpl}).  However,  our analysis on the sub-manifold is indicative of the inaccessibility of nonzero imaginary $\Sigma_{LR}$ solutions.   We thus conclude that
\be
\omega_n &<&\omega_{\rm cr} \; :\quad \Sigma_{LR}(\omega_n) \ne 0 \quad
\text{(Broken Replica Symmetry)}\nn\\
\omega_n &>& \omega_{\rm cr}\; : \quad \Sigma_{LR}(\omega_n) = 0  \quad
\text{(Replica Diagonal Solution)}
\ee
The corresponding partition function is given by
\be
F(\omega_n<\omega_{\rm cr}) &=&F_{\rm \;2\; SYK}(\omega_n),\nn\\
F(\omega_n>\omega_{\rm cr}) &=&F_{\rm\; Broken}(\omega_n).
  \ee
  The free energy of the replica diagonal solution is just the free energy of
  two decoupled SYK models \cite{Cotler:2016fpe}.

  Summing over all Matsubara frequencies we obtain the total free energy
  \be
  F &=&\sum_{\omega_{\rm cr}>\omega_n >0}
   F_{\rm\; Broken}(\omega_n)
    +\sum_{\omega_n\geq\omega_{\rm cr} }F_{\;2\;SYK}(\omega_n)\nn\\
    &=&\sum_{\omega_n>0} \log \omega_n^2+ \sum_{\omega_{\rm cr}>\omega_n >0} ( F_{\rm\; Broken} -\log \omega_n^2)
    +\sum_{\omega_n\geq \omega_{\rm cr} }(F_{\;2\;SYK}-\log \omega_n^2) 
    \ee
    The first term can be evaluated  using zeta function
    regularization:
    \be\label{eqn:zetaReg}
 \sum_{\omega_n >0} \log \omega_n^2 &=& -2\left . \frac d{ds} \left[ \frac{1}{(2\pi T)^s}
 \sum_{n>0} \frac 1{(n+1/2)^s} \right]\right|_{s=0} \nn\\
  &=& \log 2.
  \ee
  This result  gives the entropy of noninteracting
  Majorana particles.
  Our final expression for the free energy is given by
\be
  -\frac{2F}{NT} &=& \log 2+ \sum_{\omega_{\rm cr}>\omega_n >0}\left\{
 \frac {\omega_n^2}{\mjt^2-\mj^2} -1 +\log\left(\frac{\mjt^2}{\omega_n^2}\right)  \right \}
 \nn \\
&&+\sum_{\omega_n \geq \omega_{\rm cr}}\left\{-\frac{\omega_n^2}{4\mj^2} \left(1-
 \sqrt{1 + \frac{4\mj^2}{\omega_n^2}}
\right)^2+ \log\left[ \frac 14 \left(1 + \sqrt{ 1+ \frac{4\mj^2}{\omega_n^2}}\right)^2\right] \right\} .
\label{free-sd}
 \ee
 For $k=1$ we have that $\mj=0$ and  the last term vanishes.  We note that the
 free-fermion expression \eref{free-ff} agrees with \eref{free-sd} after identifying the parameters by
 \begin{equation}
\E_0^2- \y_0^2 = \mathcal{J}^2, \ (\E_0+\y_0)^2 =  \tilde{ \mathcal{J}}^2 , \ 1-\frac{\E_0}{\y_0} = \frac{2\mathcal{J}^2}{\mathcal{J}^2 -\tilde{ \mathcal{J}}^2}, \ \y_0 = \frac{\tilde{\mathcal{J}}^2 -\mathcal{J}^2}{2 \tilde{\mathcal{J}}}.
\end{equation}

\begin{figure}
\centerline{\includegraphics[width=8cm]{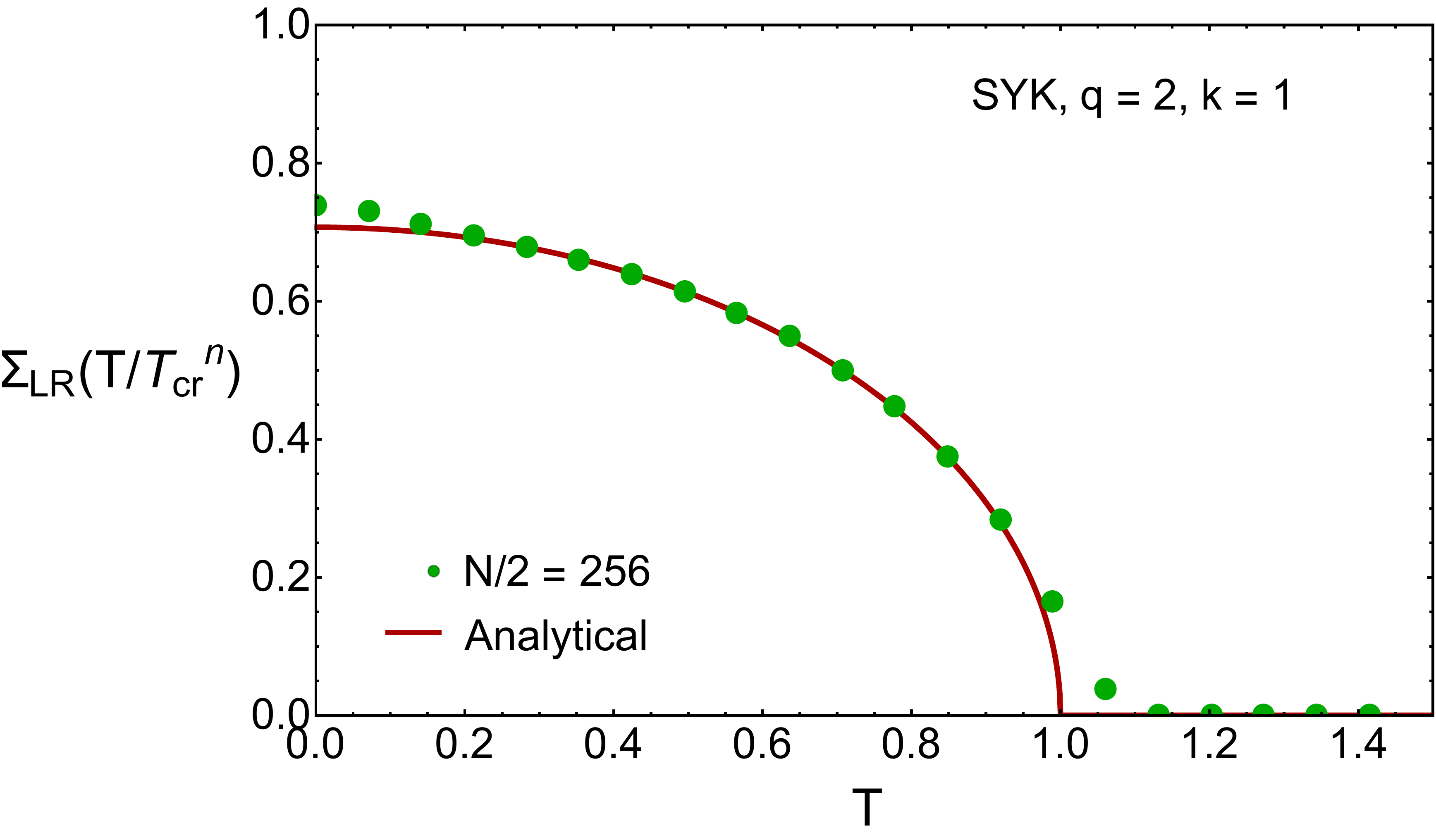}}
\caption{The order parameter $\Sigma_{LR}$ versus the temperature in units of
      the critical temperature compared to the value of $\Sigma_{LR}$ calculated
      from the eigenvalues of the hermiticized coupling matrix (see section
      \ref{sect:sigma}).}
\label{fig:cond}
\end{figure}

The order parameter of the phase transition is given by $\Sigma_{LR}$. In figure
\ref{fig:cond}
we show the analytical result \eref{sad-nt} compared with a numerical calculation
for $N/2=256$ that will be discussed in section \ref{sect:sigma}.

 The $\Sigma$ model that is obtained after integration over the $G$ variables
 resembles the usual random matrix theory $\sigma$-model. In the next section
 we will derive  basically same the $\sigma$-model directly starting from a partition function
 that is factorized into a product over Matsubara frequencies.

 \section{\boldmath Nonlinear $\sigma$-Model for $q=2$ partition function}
 \label{sect:sigma}
 
 From equation \eqref{h-ff}, we can express
the partition function of the $q=2$ Hamiltonian as
 \be
\langle Z\rangle= \left \langle \prod_{k=1}^{N/4} 2\cosh \beta \elambda_k 
  \prod_{k=1}^{N/4} 2\cosh \beta \elambda_k^*  \right \rangle,
  \ee
  where $\elambda_k$ are the eigenvalues of $i(J_{ij}+ik M_{ij})/2$ with positive real parts.  Using the Weierstrass formula
    this can rewritten in terms of
  a product over Matsubara frequencies  $\omega_n = 2\pi(n+\frac 12)/\beta$ :
\be
\langle Z \rangle&=&  2^{N/2}\left \langle \prod_{\omega_n>0} \omega_n^{-N}
\prod_{\pm\elambda_k}\left(\omega_n- 2i \elambda_k \right)
  \prod_{\pm\elambda_k} \left(\omega_n+ 2i\elambda_k^* \right) \right \rangle\nn\\
&=&  \left \langle  \prod_{\omega_n>0}
\det \left (  \omega_n + h\right)
 \det \left( \omega_n+ h^\dagger \right) \right \rangle,
 \ee
  where the product $\prod_n \omega_n $  has been evaluated to be $\sqrt{2}$ by  zeta
  function regularization (see equation \eqref{eqn:zetaReg}),  and $h$ is the $N/2 \times N/2$ skew-symmetric matrix $J+ikM$. To leading order in $1/N$
  we have 
     \be
\langle  Z\rangle =\left \langle \prod_{\omega_n>0} Z(\omega_n)\right \rangle
=  \prod_{\omega_n>0}
\langle Z(\omega_n) \rangle+O(1/N).
  \ee

  Let us first consider a one-site  SYK at frequency $\omega_n$. The
  average partition
    function is given by
    \be
\langle Z \rangle= \left \langle  \prod_{\omega_n>0}
\det \left (\omega_n+h\right)\right\rangle.
\ee
The determinant
can be expressed  \cite {Liao:2021ofk} as an integral over Grassmann variables $\phi$ and $\phi^*$:
\be
\vev{Z(\omega_n)}=\left \langle \int \prod_i d\phi_i d\phi_i^* e^{\sum_{ij}\phi^*_i( \omega_n\delta_{ij}+h_{ij}) \phi_j}
\right \rangle
\ee
We recall that $\vev{J_{ij}^2} =\vev{M_{ij}^2} ={ \cal J}^2/(1-k^2)$, so the Gaussian average  over $h$ can be performed by a cumulant expansion
resulting in
\be
\vev{Z(\omega_n)}&=&
\int \prod_i d\phi_i d\phi_i^* e^{\omega_n (\phi^* \cdot \phi)
  +\frac 1{N} {\cal J}^2\sum_{i<j} (\phi_i^* \phi_j
  -\phi_j^* \phi_i)^2 
}
\nn\\
&=& \int\prod_i d\phi_i d\phi_i^* e^{ \omega_n (\phi^*\cdot \phi)
  -\frac 1{N}  {\cal J}^2\sum_{ij} \phi_i^* \phi_j
  \phi_j^* \phi_i
}\nn \\
&=& \int\prod_i d\phi_i d\phi_i^* e^{  \omega_n (\phi^*\cdot \phi)
 +\frac 1{N}  {\cal J}^2(\phi^*\cdot \phi)^2
}
\ee
where we have used $\phi_i^2 =0$ due to their Grassmannian nature. 
Using the Hubbard-Stratonovich transformation
  \be
  e^{ \alpha^2 A^2/2} = \frac 1{\sqrt{2\pi\alpha^2}}
    \int d\Sigma e^{-\frac {\Sigma^2}{2\alpha^2}-\Sigma A} 
\label{hs}
    \ee
    we obtain
    \be
\vev{Z(\omega_n)}&= &\int\prod_i d\phi_i d\phi_i^* d\Sigma
e^{ \omega_n(\phi^*\cdot \phi)
  -\frac  N 4 \frac{ \Sigma^2}{\mj^2} - \Sigma  (\phi^* \cdot \phi)}.
\nn\\
&=&\int d \Sigma \ (  \omega_n -\Sigma)^{N/2}
e^{  -\frac N4 \frac{\Sigma^2}{ \mj^2} }.
\ee
The saddle point equation is 
\be
\Sigma^2 - \omega_n \Sigma -\mj^2 =0.
\ee
The dominant solution is given by  (we have $\omega_n>0$ to start with)
\be
\Sigma = \frac {\omega_n}2 -\frac 12 \sqrt{\omega^2_n +4\mj^2}.
\ee
with leads to the one-site free energy  
\be
-\frac {F(\omega_n)}{T N/2} = \log\left (\frac{\omega_n}2 + \frac {1} 2
\sqrt{\omega_n^2+4\mj^2}\right )  -\frac 1{2\mj^2} \left (\frac{\omega_n}2  -\frac 12 \sqrt{\omega_n^2 +4\mj^2} \right )^2.
     \ee
     This is exactly half of the  the Schwinger-Dyson result for the two-site model using only the replica-symmetric saddles.  The total free energy is given
     by
     \be
     -\frac F{T N/2} =\sum_{\omega_n >0} -\frac {F_{\;2\;SYK}(\omega_n)}{T N} .
     \ee
     This shows explicitly that the one-site annealed partition function
     gives the quenched result for the high-temperature phase of the of
     the nonhermitian SYK model as is the case for the $q=4$ SYK model.
     For $T<T_c$ the replica limit of the one-site partition function
 fails to give the quenched result. In order to get the correct result we have to take the
     replica limit of the partition function and the conjugate partition function,
     which is well-known from the $\sigma$-model
     formulation of nonhermitian random matrix theories  \cite{girko2012theory,efetov1997directed,feinberg1997non} and QCD at nonzero chemical potential
\cite{stephanov1996,Janik:1996va,Janik:1996xm,splittorff:2003cu}.

Next we consider the two-site  non-Hermitian model and also assume that the ensemble
average factorizes in the large $N$ limit so that we can evaluate the
partition function for a single frequency
 \be
\langle Z(\omega_n)\rangle&=&  \left \langle  
\det \big (\omega_n+h^{}\big )
 \det \big (\omega_n+ h^\dagger \big ) \right \rangle,
 \nn\\
 &=&  \int \left(\prod_{i}^{N/2}d\phi^i_L d\phi^{i*}_L d\phi^i_R d\phi^{i*}_R\right) e^{\phi_L^* \cdot(\omega_n + h) \cdot\phi_L +
    \phi_R^*\cdot(\omega_n + h^\dagger)\cdot \phi_R}. 
 \ee
 We can again average over $h$ by a cumulant expansion using that
 \be
 \langle h_{ij}^2 \rangle &=&\langle h_{ij}^2 \rangle = \frac {{\cal J}^2}{ N/2} ,\nn\\
 \langle h_{ij} h_{ij}^* \rangle &=& \frac{\widetilde {\cal J}^2}{ N/2}.
 \ee
 This results in the quartic action 
 \be
 -S_4 = \frac{ \mj^2}{N} \left [ (\phi_R^*\cdot \phi_R)^2 + (\phi_L^*\cdot \phi_L)^2 \right ] + \frac{2 \mjt^2}{N} \left[ (\phi_L^* \cdot \phi_R^*) (\phi_L \cdot \phi_R) +  (\phi_L^*\cdot \phi_R) (\phi_L\cdot \phi_R^*)\right ].
 \ee
  This action is invariant under
 \be
 \phi_L \to e^{i\varphi} \phi_L, \qquad  \phi_R \to e^{i\psi} \phi_R.
\label{symm}
 \ee
 In addition to the Hubbard-Stratonovich transformation \eref{hs} we use the
 identity
 \be
 e^{\alpha^2 A A^* } \sim e^{-\frac {\Sigma \Sigma^*}{\alpha^2}- \Sigma A^* -\Sigma^* A}
 \ee
 to decouple the quartic terms. This results in
 \be
 Z(\omega_n) =\int d\Sigma d\rho e^{ -\frac N{4\mj^2} \left[ \Sigma_{RR}^2 +\Sigma_{LL}^2\right ]
      -\frac {N}{2\mjt^2}
   \left[ \Sigma^*_{LR}\Sigma_{LR}  +\rho_{LR}^{}\rho_{LR}^*\right ]}
 I^{N/2}(\{\Sigma,\rho\}),
 \ee
 where $d \Sigma = d\Sigma_{LL}d\Sigma_{RR} d\Sigma_{LR} d\Sigma^*_{LR}$
 and $d\rho= d\rho_{LR} d\rho^*_{LR} $ and we have used that the integral
 over the $\phi^i_a$ variables factorizes into a product over $I$. Each
 of these factors is equal to
\be
I&=&	\int d\phi_L d\phi^{*}_L d\phi_R d\phi^{*}_R\exp[(\omega_n -\Sigma_{LL})\phi_L^* \phi_L +(\omega_n -\Sigma_{RR})\phi_R^*  \phi_R \nn\\
 && \hspace*{4cm} -\Sigma_{LR}\phi_L^*\phi_R - \Sigma_{LR}^*\phi_L\phi_R^*
- \rho_{LR}\phi_L^*\phi_R^*-\rho_{LR}^*\phi_L\phi_R],
\ee
where the $\phi_a$ no longer carry an index $i$. This integral 
 can be evaluated by  simply  expanding the exponential in the integrand and collecting the terms that are proportional to $\phi_L \phi^{*}_L\phi_R \phi^{*}_R$. We get 
\be 
I =  (\omega_n-\Sigma_{LL})(\omega_n-\Sigma_{RR})+\Sigma_{LR}^*\Sigma_{LR}
  -\rho_{LR}^*\rho_{LR}.
\ee 
The saddle points can be grouped in to the following three classes:
\begin{align}
&\rho_{LR} =0,\quad \Sigma_{LR}=0 , \quad \Sigma_{LL}=\Sigma_{RR} = \frac{\omega_n}{2}-\frac{1}{2}\sign(\omega_n)\sqrt{\omega_n^2+4\mj^2},\\
&\rho_{LR} = 0,  \quad |\Sigma_{LR}|^2= \mjt^2\left(1-\frac{\omega_n^2}{\omega_{\rm cr}^2}\right), \quad \Sigma_{LL}=\Sigma_{RR} = -\frac{\mj^2 \omega_n}{\mjt^2-\mj^2}, \\
&|\rho_{LR}|^2 =\mjt^2\left[1+\frac{\mjt^2 \omega_n^2}{(\mjt^2+\mj^2)^2}\right], \quad \Sigma_{LR}=0,\quad \Sigma_{LL}=\Sigma_{RR} = -\frac{\mj^2 \omega_n}{\mjt^2+\mj^2}.
\end{align}
Now the first two saddles are simply the ones we found in the SD equations in section \ref{sec:sd},  and if the third saddle can be discarded they would reproduce the same saddle-point analysis, which we recall here: although the second saddle (symmetry-breaking saddle) has a dominant action (larger $-S_E$) for all values of $\omega_n$,   thimble analysis requires us to pick the second saddle only for $\omega_n< \omega_{\rm cr}$ and for  $\omega_n> \omega_{\rm cr}$ we must pick the first saddle (unbroken saddle).  Let us now  show why the third saddle can be discarded:  the third saddle's action ($-S_E$) is smaller than that of the second saddle for all $\omega_n$,  so we do not need to worry about it for $\omega_n <\omega_{\rm cr}$;  its action is smaller than that of the first saddle for $\omega_n >\omega_{\rm cr}$ (its action can be larger than that of the first saddle only for $\omega_n<\omega_{\rm cr}$), hence we do not need to worry about the third saddle for $\omega_n >\omega_{\rm cr}$ either.    Thus,  the third saddle drops out of our consideration for all values of $\omega_n$  and we reproduce the same free energy found in section  \ref{sec:sd}.

The order parameter of the phase transition of the coupled SYK model is given
by the expectation value of 
$\Sigma_{LR}$. This is equal to
\be
\langle\Sigma_{LR}  \rangle =-\frac {2\mjt^2}{N}
\langle \phi_L \cdot \phi_R^*  \rangle
\ee
This
is the chiral condensate corresponding
 to the spectral density of
 \be
    {\cal H} = \bmat 0 & h + \omega_n \\
    h^\dagger +\omega_n & 0 \emat .
    \ee
    It is given by the Banks-Casher formula \cite{Banks:1979yr}
    \be
    \Sigma_{LR} =- \lim_{\epsilon\to 0}\lim_{N\to \infty}\frac {2\mjt^2}{N} \Tr \frac 1{{\cal H} +i\epsilon}
=   \lim_{N\to \infty}  \frac {\pi \rho_{\cal H}(0)}N.
      \ee
      The critical temperature is determined by the value of
      $\omega_n$ at which a gap opens and
      the spectrum becomes
      gapped  for $T>T_c$. Another interpretation of the critical temperature
      is that
      $T_c$ is the point at which $\omega_n$ enters the spectral support
      of $h$. The spectral density of ${\cal H}$ can be obtained analytically and
        follows from the solution of a cubic
      equation \cite{Jackson:1995nf,Jackson:1996xt}. The phase transition
      is a typical Landau-Ginsberg phase transition with mean field critical
      exponents.

      The partition function $Z(\omega_n) $ with $h$ replaced by a complex
      matrix was first introduced  as a random matrix model
      for chiral symmetry breaking in QCD \cite{Jackson:1995nf} and further analytical results
      were obtained in a subsequent paper \cite{Jackson:1996xt}.

  \section{Outlook and conclusions}\label{sec:outlook}
 In conclusion, the free energy of the integrable $q=2$ SYK model is qualitatively
 different from the $q >2$ case. Not only is the order of the transition is different,
 but also  there is an infinite series of transitions while for $q>2$
 there is only one. This goes back to the factorization of partition function
 into a product over Matsubara frequencies. Each of
 the factors undergoes a phase transition
 from a replica symmetric solution to a solution with broken replica symmetry
 with a critical temperature that depends on the Matsubara frequency.
 For the full partition function this results into an infinite sequence of
 phase transitions.
 
 We have calculated the quenched free energy in two structurally different ways. First,
 a quenched calculation based on the free-fermion description of the $q=2$
 SYK model, and second, an annealed calculation based on the solution of the
 Schwinger-Dyson equations in the $\Sigma G$ formulation of the SYK model
using the replica trick.
 The two methods give the same result which shows that despite the nonhermiticity
 of the model, the replica limit gives the correct result provided that
 the starting point is the product of the one-site partition function
 and its complex conjugate before averaging. On the other hand, in the
 quenched free-fermion calculation, we did start from the one-site partition
 function without having to include the conjugate partition function.  The
 reason this gives the correct result is the factorization of the partition function in a product over
 single particle energies.

 The $q=4$ nonhermitian SYK model behaves quite differently. It is not a Fermi liquid and the usual free-fermion description is invalid which is most notable
 in the zero temperature entropy which is extensive \cite{Sachdev:2001}. The nonhermitian
 two-site $q=4$ SYK model
 has a single first order phase transition which separates a low-temperature
 phase from a high-temperature phase.
 The high-temperature phase is entropy dominated while the low-temperature
 phase is energy dominated. The free energy in the high-temperature phase
 follows from the many-body eigenvalue density and is entirely determined
 by the one-site partition function. This is also the case for the $q=2$ SYK model.
 In
 the low-temperature phase, the replica limit of the one-site partition
 function breaks down and the quenched one-site partition function is given
 by the replica limit of the one-site partition function and its complex
 conjugate.
 In terms of the many-body spectral density, the two-point spectral correlation
 function determines the free-energy of the low-temperature phase. For this
 reason there is large difference between the $q=4$ case and the $q=2$ case.
 The dynamics of the $q=4$ partition function is chaotic with universal
 eigenvalue
 correlations given by the Ginibre model, which as a consequence gives rise to a
 temperature-independent free energy in the low-temperature phase.
 On the other hand, the $q=2$ SYK
 model is integrable with mostly but not entirely uncorrelated eigenvalues.
We can distinguish two contributions from the two-point correlation function.
 One contribution is due to self-correlations, and the second one is
 due to the many-body correlation resulting from the fact that the
 $2^{N/2}$ many-body
 eigenvalues are determined by $N/2$ single-particle energies. It is simple
 to evaluate the contribution from the self-correlations but this only
 reproduces the free-energy at zero temperature and is smooth as a function of
 the temperature. This implies that the infinite series of second-order
 phase transitions are due to  correlations of the many-body
 eigenvalues.

 The phase transitions of the $q=2$ nonhermitian SYK model can also be understood
 in terms of the spectral properties of the coupling matrix.
 Using a random matrix theory like $\sigma$-model calculation we  have related
 the order parameter of the phase transition $\Sigma_{LR}$ to the formation of a gap
 of the hermiticized two-site Hamiltonian. In terms of the one-site Hamiltonian,
 this is the point where the Matsubara frequency enters the support of the
 spectrum of $H_L$. The starting point of the $\sigma$-model calculation
 is closely related to a random matrix model for the chiral phase transition
 in QCD where $\Sigma_{LR}$ plays the role of the chiral condensate.

 A natural question is whether the free energy of the $q=2$ SYK
 model can also be understood in terms of the many-body spectral density
 and the many-body spectral correlations. The cancellations that are
 responsible for the high-temperature phase of the $q=4$ model are still
 at work for $q=2$. For example, at high temperatures and for maximum nonhermiticity the free energy
 of the $q=2$ SYK model and the $q=4$ SYK model is the same ($- \frac T2 \log 2$ per particle).
From the solutions of the SD equations it is clear  that the
two-point correlation function determines the low-temperature phase. In particular, the transitions observed in the
 low-temperature phase are due to   the coupling
 between left and right sites but not by the dynamics within each of the sites.

 In conclusion, the nature of the quantum dynamics plays a role for the replica symmetry breaking mechanism which also induces phase transitions  for the
 $q=2$ nonhermitian SYK model.
However, while the replica dynamics of quantum chaotic systems is universal, there is a broad variety of dynamical behavior associated with integrable systems,
and we cannot conclude that the behavior we have observed for the
nonhermitian $q=2$ SYK model is generic.

\acknowledgments{JV would like to acknowledge Freeman Dyson for responding
  with a hand written letter when I applied for a postdoc position in 1981.  This response has encouraged my work on RMT throughout the years.  YJ would like to thank Freeman Dyson for a conversation that happened in 2013 in Singapore,   which has a lasting impact on YJ's academic personality. Antonio Garc\'ia-Garc\'ia is
  thanked for collaboration in early stages of this project and providing
  eigenvalues of the $q=2$ nonhermitian SYK Hamiltonian. Gernot Akemann and
  Yuya Tanizaki are thanked for pointing out and explaining references  \cite{hastings2001, hamazaki2020, akemann2022spacing,Kanazawa:2014qma,Tanizaki:2015gpl}.
 	YJ and JV  acknowledge partial support from U.S. DOE Grant
 	No. DE-FAG-88FR40388. YJ is also partly funded
by an Israel Science Foundation center for excellence grant (grant number 2289/18), by grant number 2018068 from the United States-Israel Binational Science Foundation (BSF), by the Minerva foundation with funding from the Federal German Ministry for Education and Research,  by the German Research Foundation through a German-Israeli Project Cooperation (DIP) grant ``Holography and the Swampland'' and by Koshland postdoctoral fellowship.  DR acknowledges support from the Korea Institute of Basic Science (IBS-R024-Y2 and IBS-R024-D1).}

\appendix

\section{Free-fermion representation of the $q=2$ SYK model}\label{a:yiyangfreefermions}
We write the $q=2$ one-site Hamiltonian as 
 \begin{equation}\label{eqn:nonHermq=2Hami}
 H_L =\frac{1}{2} \sum_{ij} W_{ij} \gamma_i\gamma_j = \frac{1}{2} \vec{\gamma}^TW \vec{\gamma}.
 \end{equation}
 where gamma matrices $\vec{\gamma} = (\gamma_1,\gamma_2,\ldots,\gamma_{M})$ with even $M$ and $W$ is a random antisymmetric complex matrix. In the main text we have the convention
 that $M=N/2$.
  By matching with definitions \eqref{eq:sykq2} and \eqref{eqn:majoranaNormalization}, we know 
 \begin{equation}
 W = \frac{1}{2} i (J_{ij}+ikM_{ij}).
 \end{equation}
 In the Hermitian SYK model ($k=0$),    $W$ is an antisymmetric Hermitian matrix, and the Hamiltonian can be transformed into a  fermion-filling form thanks to the fact that an $M$-dimensional antisymmetric Hermitian matrix has the normal form 
 \begin{equation}\label{eqn:normalFormAntiSymHermi}
 O \begin{pmatrix}
 i \elambda_1 \sigma_2 & 0 & \cdots &0 \\
 0 &  i \elambda_2 \sigma_2 & \cdots & \vdots \\
 \vdots & \cdots  &\cdots & \vdots\\
 0  & \cdots  &\cdots & i \elambda_{\frac{M}2} \sigma_2
 \end{pmatrix} O^T,
 \end{equation}
 where $O$ is a real orthogonal matrix. 
 Using a new basis for the $\gamma$ matrices defined by $O^T \vec{\gamma}$, one can easily write the Hermitian Hamiltonian in a fermion-filling form.
Although a generic complex antisymmetric matrix does not have a normal form of equation \eqref{eqn:normalFormAntiSymHermi},  there exists a parallel of it which allows us to write the non-Hermitian  Hamiltonian \eqref{eqn:nonHermq=2Hami} in a modified fermion-filling form. We will demonstrate this now. 

We consider a diagonalizable complex antisymmetric matrix $W$ with all eigenvalues being nonzero,  which is almost always the case for our ensemble.   The eigenvalues of $W$ come in opposite pairs $\pm \elambda$, so we can diagonalize $W$ as 
\begin{equation}\label{eqn:Wdiag}
W = S \Lambda S^{-1},
\end{equation}
where 
\begin{equation}
\Lambda = \begin{pmatrix}
  \elambda_1 \sigma_3 & 0 & \cdots &0 \\
 0 &   \elambda_2 \sigma_3 & \cdots & \vdots \\
 \vdots & \cdots  &\cdots & \vdots\\
 0  & \cdots  &\cdots &  \elambda_{\frac{M}2} \sigma_3
 \end{pmatrix} .
\end{equation}
The column vectors of $S$ are the eigenvectors of $W$,  namely
\begin{equation}
S= \begin{pmatrix}
  \vline &  \vline & \cdots &  \vline  &  \vline\\
   \vline &   \vline& \cdots &   \vline   &  \vline\\
 v_1 & v_2  &\cdots & v_{M-1}& v_{M}\\
  \vline &   \vline& \cdots &   \vline  &  \vline \\
    \vline &   \vline& \cdots &   \vline  &  \vline\\
   \end{pmatrix},
\end{equation}
where 
\begin{equation}
W v_{2k-1} = \elambda_{k}v_{2k-1}, \quad W v_{2k} =- \elambda_{k}v_{2k}.
\end{equation}
To completely fix the sign convention, we choose $\elambda_k\ (k=1,\ldots,M/2)$ to have a positive real part.
Because $W^T = -W$, we have 
\begin{equation}
0= v_i^T \left(W+W^T\right)v_j=  (\elambda_i+\elambda_j) v_i^T v_j 
\end{equation}
which implies
\begin{equation}
 v_i^T v_j  = 0  \text{ unless } \{i,j\} = \{2k-1,2k\}.
\end{equation}
So if we scale the eigenvectors to redefine $S$ as 
\begin{equation}
S= \begin{pmatrix}
  \vline &  \vline & \cdots &  \vline  &  \vline\\
   \vline &   \vline& \cdots &   \vline   &  \vline\\
 \frac{v_1}{\sqrt{v_1^T v_2}} &  \frac{v_2}{\sqrt{v_1^T v_2}}  &\cdots & \frac{v_{D-1}}{\sqrt{v_{M-1}^T v_M}}& \frac {v_{M}} {\sqrt{v_{M-1}^T v_M}}\\
  \vline &   \vline& \cdots &   \vline  &  \vline \\
    \vline &   \vline& \cdots &   \vline  &  \vline\\
   \end{pmatrix},
\end{equation}
we can easily see
\begin{equation}
S^T S = \begin{pmatrix}
\sigma_1 & 0 & \cdots &0 \\
 0 & \sigma_1 & \cdots & \vdots \\
 \vdots & \cdots  &\cdots & \vdots\\
 0  & \cdots  &\cdots & \sigma_1
 \end{pmatrix} \equiv \Sigma_1,
\end{equation}
and hence $S^{-1} =\Sigma_1 S^T$ (note in general $S^T S\neq S S^T$). Substituting this into equation \eqref{eqn:Wdiag}, we obtain 
\begin{equation}
W = S \Lambda \Sigma_1 S^T = 
S \begin{pmatrix}
 i \elambda_1 \sigma_2 & 0 & \cdots &0 \\
 0 &  i \elambda_2 \sigma_2 & \cdots & \vdots \\
 \vdots & \cdots  &\cdots & \vdots\\
 0  & \cdots  &\cdots & i \elambda_{\frac{M}2} \sigma_2
 \end{pmatrix} S^T.
\end{equation}
We thus arrive at a normal form for complex antisymmetric matrices rather similar to that of the Hermitian antisymmetric ones \eqref{eqn:normalFormAntiSymHermi}, with the difference that $S$ is not orthogonal but satisfies $S^T S =\Sigma_1$.   Now we define new set of operators $\{\tilde c_k, c_k | k=1,2\ldots M/2\}$ by 
\begin{equation}
\tilde c_k = \frac{1}{\sqrt{2}}\left(S^T \vec{\gamma}\right)_{2k-1}, \quad c_k = \frac{1}{\sqrt{2}}\left(S^T \vec{\gamma}\right)_{2k}. 
\end{equation}
From the anti-commutation relation of $\gamma$ matrices we derive
\begin{equation}
\frac{1}{2}\left\{\left(S^T \vec{\gamma}\right)_m,  \left(S^T \vec{\gamma}\right)_n\right\} = (\Sigma_1)_{mn}.
\end{equation}
This in particular implies 
\begin{equation}
\tilde{c}_k^2=0, \quad,  c_k^2 = 0,  \quad \{c_k,\tilde c_l\} = \delta_{kl}.
\end{equation}
Note this is just the algebra for the ladder operators of $M/2$ spinless fermions, except that $c_k$ and $\tilde c_k$ are not related by a Hermitian conjugation.  In terms of these ladder operators, the Hamiltonian \eqref{eqn:nonHermq=2Hami} becomes
\begin{equation}
H = \sum_{k=1}^{M/2} \elambda_k (2\tilde c_k c_k -1).
\end{equation} 
Just as the in Hermitian case,  we have 
\begin{equation}
[H, \tilde c_k] =\elambda_k \tilde c_k, \quad [H,  c_k] =-\elambda_k c_k.
\end{equation}
Hence we conclude the many-body energies of $H$ are given by the filling of $M/2$ free fermions into the particle-hole symmetric levels of $W$: each fermion either occupies a ``particle'' level with energy $\elambda_k$, or occupies a ''hole'' level with energy $-\elambda_k$.    However since the raising and lowering operators are not Hermitian conjugate to each other, the eigenstates are not necessarily orthogonal to each other (at least not with the original inner product $\langle x, y\rangle \equiv x^\dagger y$), just as one would expect for a non-Hermitian Hamiltonian.

\section{Free energy from the free-fermion representation for $k< 1$}
\label{app:b}
Based on the free-fermion representation of the $q =2$ SYK model of appendix \ref{a:yiyangfreefermions}, we proceed to the explicit analytical calculation of the free energy. The simpler spherical case $k = 1$ was already discussed in the
main text. The free energy for $k < 1$ can be derived along the same lines
which is the purpose of this appendix.

For $k<1$ ($\y_0 <\E_0$),    the large-$N$ single-particle spectral density becomes a constant inside an elliptical disk as in equation \eqref{eqn:specDenEllipse}.  The elliptical disk can be parameterized by 
\begin{equation}\label{eqn:paramEllipse}
z = \E_0 r \cos \phi + i \y_0 r \sin \phi, \quad r\in [0,1],\; \phi \in [0,2\pi].
\end{equation}
We write 
\begin{equation}\label{eqn:freeEnIntegralEllipse}
-\frac{F_L}{(N/4)T} =\int_0^{1} r I_E(r) dr ,
\end{equation}
where 
\begin{equation}
I_E(r) = \frac{1}{\pi}\int_0^{2\pi}  \log \left(2 \cosh \frac{\E_0 r \cos \phi + i \y_0 r \sin \phi}{T}\right)d \phi.
\end{equation}
The subscript $E$ denotes ``Ellipse''.  On its face,  we cannot interpret $I_E(r)$ as a complex contour integral like equation \eqref{eqn:contourInt},  because $dz/(iz) \neq d\phi$ with the elliptical parameterization \eqref{eqn:paramEllipse}.    We can overcome this by considering the following conformal (Joukowski) transformation : 
\begin{equation}
z = a u + \frac{b}{u},  
\end{equation}
where
\begin{equation}
\quad a = \frac{\E_0+\y_0}{2}r, \quad b= \frac{\E_0-\y_0}{2}r.
\end{equation}
In terms $u$, the ellipse in equation \eqref{eqn:paramEllipse} at any given $r$ becomes a unit circle:
\begin{equation}
u = e^{i\phi}
\end{equation}
We stress that the $\phi$ here is the same $\phi$ as in the parameterization \eqref{eqn:paramEllipse}.  Now we can write 
\begin{equation}
\begin{split}
I_E(r) =&\frac{1}{\pi i}\oint_{S^1} \frac{du}{ u}  \log \left(2 \cosh  \frac{ au +b/u}{T}\right)\\
= & 2\log 2 +\sum_{n=0}^\infty  \frac{1}{\pi i} \oint_{S^1} \frac{du}{u}  \log \left(1+ \frac{4(au+b/u)^2}{\omega_n^2}\right)\\
\equiv & 2 \log 2 +\sum_{n=0}^\infty I_{nE}(r),
\end{split}
\end{equation}
where to obtain second equality we applied Weierstrass factorization just as we did in the circular case and the third equality simply defines $I_{nE}$. Notice $S^1$ denotes the unit circle and the $r$-dependence of the integral comes from the $r$-dependence of $a$ and $b$.

To analyze the cut and pole structures of the integral $I_{nE}$, we rewrite its integrand as
\begin{equation}
\begin{split}
\frac{1}{u}\log \left(1+ \frac{4(au+b/u)^2}{\omega_n^2}\right) = \frac{1}{u}\log\left[ \frac{4a^2}{\omega_n^2 u^2}(u - u_{1+})(u -u_{1-}) (u-u_{2+})(u-u_{2-})\right],
\end{split}
\end{equation}
where $u_{1\pm}, u_{2\pm}$ are the four roots of the equation
\begin{equation}
1+ \frac{ 4(a u +b/u)^2}{\omega_n^2} = 0,
\end{equation}
namely 
\begin{equation}
u_{1\pm} = \pm i \frac{\sqrt{\omega_n^2/4 +4ab}-\omega_n/2}{2a}, \quad u_{2\pm} = \pm i \frac{\sqrt{\omega_n^2/4 +4ab}+\omega_n/2}{2a}.
\end{equation}
Note that $4ab =(\E_0^2 -\y_0^2)r^2>0$,  and since $a>b$,  $u_{1\pm}$ are always inside the unit circle. We also note that for $u= e^{i\phi}$
\begin{equation}
\oint_{S^1} du \frac{\log u}{ u} =0.
\end{equation}
Hence 
\begin{equation}
I_{nE} = \frac{1}{\pi i}\oint_{S^1} \frac{du}{ u}  \log\left[ \frac{4a^2}{\omega_n^2}(u - u_{1+})(u -u_{1-}) (u-u_{2+})(u-u_{2-})\right].
\end{equation}
The integrand of $I_{nE}$ has one pole at the origin with the residue 
\begin{equation}
\log\left[ \frac{4a^2}{\omega_n^2}u_{1+}u_{1-} u_{2+}u_{2-}\right] = \log \left( \frac{4b^2}{\omega_n^2} \right).
\end{equation}
The integrand of $I_{nE}$ has four branch cuts emanating from $u_{1 \pm}, u_{2\pm}$ horizontally to negative infinity.  The $u_{1\pm}$ cuts always intersect with the unit circle,  whereas   $u_{2\pm}$ may or may not intersect with the unit circle depending on the values of $a$ and $b$.  To summarize, the $u_{2\pm}$ cuts contribute to $I_{nE}$ only if $|u_{2\pm}|<1$ (which is to say $r > \omega_n/2\y_0$),  in much the same way as the $\pm i \omega_n/2\y_0$ cuts contribute to $I_n$ in the circular case; what is new with the elliptical case are the $u_{1\pm}$ cuts and the pole at the origin,  which always contribute to $I_{nE}$ regardless the value of $r$.  Recycling the calculation done in the circular case, we obtain 
\begin{equation}
I_{nE} (r) =  2 \log \left[\frac{1}{4}\left(1 + \sqrt{1+\frac{4(\E_0^2-\y_0^2)r^2}{\omega_n^2}}\right)^2 \right] 
\end{equation}
if $r<\omega_n/2\y_0$. which is the sum of one pole and two cut contributions.  And 
\begin{equation}
I_{nE} (r) =   2 \log \left[\frac{1}{4}\left(1 + \sqrt{1+\frac{4(\E_0^2-\y_0^2)r^2}{\omega_n^2}}\right)^2 \right]+2 \log \left(\frac{2(\E_0+\y_0)r}{\omega_n+\sqrt{\omega_n^2 +4(\E_0^2-\y_0^2)r^2}}\right)
\end{equation}
if $r>\omega_n/2\y_0$, which is the sum of one pole and four cut contributions.  With these results, we arrive at 
\begin{equation}
\begin{split}
-\frac{F_L}{(N/4)T}  =& \log 2 + \sum_{\omega_n >0} \int_0^1 r I_{nE}(r) dr\\
& = \log2 +  \sum_{n=0}^\infty \left\{ \log \left[\frac{1}{4}\left(1 + \sqrt{1+\frac{\E_0^2-\y_0^2}{\omega_n^2/4}}\right)^2 \right] -\frac{\omega_n^2/4}{\E_0^2-\y_0^2}\left(1-\sqrt{1+\frac{\E_0^2-\y_0^2}{\omega_n^2/4}}\right)^2     \right\}\\
& \qquad - \sum_{0<\omega_n <2\y_0}
\left\{ \frac{2\omega_n^2/4}{\E_0^2-\y_0^2} \sqrt{1+ \frac{\E_0^2-\y_0^2}{\omega_n^2/4}} +\log \left[\frac{\omega_n^2/4}{(\E_0+\y_0)^2}\left(1 + \sqrt{1+\frac{\E_0^2-\y_0^2}{\omega_n^2/4}}\right)^2 \right] \right\} \\
& \qquad + \sum_{0<\omega_n<2 \y_0}
\frac{2\omega_n^2 \E_0/4}{(\E_0^2-\y_0^2)\y_0} \\
=&\log 2  + \sum_{\omega_n > 2 \y_0}
 \left\{ \log \left[\frac{1}{4}\left(1 + \sqrt{1+\frac{\E_0^2-\y_0^2}{\omega_n^2/4}}\right)^2 \right] -\frac{\omega_n^2/4}{\E_0^2-\y_0^2}\left(1-\sqrt{1+\frac{\E_0^2-\y_0^2}{\omega_n^2/4}}\right)^2     \right\}\\
&\qquad  + \sum_{0<\omega_n <2 \y_0}
\left\{ \log \frac{(\E_0+\y_0)^2}{\omega_n^2}  - 1 - \frac{\omega_n^2(1-\E_0/\y_0)}{2(\E_0^2 - \y_0^2)}\right\}.
\end{split}
\end{equation}
The result on the last line exactly matches with the SD calculation provided that
\begin{equation}
\E_0^2- \y_0^2 = \mathcal{J}^2, \ (\E_0+\y_0)^2 =  \tilde{ \mathcal{J}}^2 , \ 1-\frac{\E_0}{\y_0} = \frac{2\mathcal{J}^2}{\mathcal{J}^2 -\tilde{ \mathcal{J}}^2}, \ \y_0 = \frac{\tilde{\mathcal{J}}^2 -\mathcal{J}^2}{2 \tilde{\mathcal{J}}}.
\end{equation}

\bibliographystyle{JHEP}
\bibliography{librarynh}

\providecommand{\href}[2]{#2}\begingroup\raggedright\begin{thebibliography}{10}

\bibitem{dyson1996selected}
F.J.~Dyson et~al., \emph{Selected papers of Freeman Dyson with commentary},
  vol.~5, American Mathematical Soc. (1996).

\bibitem{Dyson:1972tm}
F.J.~Dyson, \emph{{A class of matrix ensembles}},
  \href{https://doi.org/10.1063/1.1665857}{\emph{J. Math. Phys.} {\bfseries 13}
  (1972) 90}.

\bibitem{bethe1936}
H.A.~Bethe, \emph{An attempt to calculate the number of energy levels of a
  heavy nucleus}, \href{https://doi.org/10.1103/PhysRev.50.332}{\emph{Phys.
  Rev.} {\bfseries 50} (1936) 332}.

\bibitem{french1970}
J.~French and S.~Wong, \emph{Validity of random matrix theories for
  many-particle systems},
  \href{https://doi.org/http://dx.doi.org/10.1016/0370-2693(70)90213-3}{\emph{Physics
  Letters B} {\bfseries 33} (1970) 449 }.

\bibitem{french1971}
J.~French and S.~Wong, \emph{Some random-matrix level and spacing distributions
  for fixed-particle-rank interactions},
  \href{https://doi.org/http://dx.doi.org/10.1016/0370-2693(71)90424-2}{\emph{Physics
  Letters B} {\bfseries 35} (1971) 5 }.

\bibitem{bohigas1971}
O.~Bohigas and J.~Flores, \emph{Two-body random hamiltonian and level density},
  \href{https://doi.org/http://dx.doi.org/10.1016/0370-2693(71)90598-3}{\emph{Physics
  Letters B} {\bfseries 34} (1971) 261 }.

\bibitem{bohigas1971a}
O.~Bohigas and J.~Flores, \emph{Spacing and individual eigenvalue distributions
  of two-body random hamiltonians},
  \href{https://doi.org/http://dx.doi.org/10.1016/0370-2693(71)90399-6}{\emph{Physics
  Letters B} {\bfseries 35} (1971) 383 }.

\bibitem{mon1975}
K.~Mon and J.~French, \emph{Statistical properties of many-particle spectra},
  \href{https://doi.org/http://dx.doi.org/10.1016/0003-4916(75)90045-7}{\emph{Annals
  of Physics} {\bfseries 95} (1975) 90 }.

\bibitem{kitaev2015}
A.~Kitaev, \emph{A simple model of quantum holography}, {\emph{KITP strings
  seminar and Entanglement 2015 program, 12 February, 7 April and 27 May 2015,
  {\tt \url{http://online.kitp.ucsb.edu/online/entangled15/}}} (2015) }.

\bibitem{maldacena2016}
J.~Maldacena and D.~Stanford, \emph{{Remarks on the Sachdev-Ye-Kitaev model}},
  \href{https://doi.org/10.1103/PhysRevD.94.106002}{\emph{Phys. Rev. D}
  {\bfseries 94} (2016) 106002}
  [\href{https://arxiv.org/abs/1604.07818}{{\ttfamily 1604.07818}}].

\bibitem{sachdev1993}
S.~Sachdev and J.~Ye, \emph{Gapless spin-fluid ground state in a random quantum
  heisenberg magnet},
  \href{https://doi.org/10.1103/PhysRevLett.70.3339}{\emph{Phys. Rev. Lett.}
  {\bfseries 70} (1993) 3339}.

\bibitem{sachdev2015}
S.~Sachdev, \emph{Bekenstein-hawking entropy and strange metals},
  \href{https://doi.org/10.1103/PhysRevX.5.041025}{\emph{Phys. Rev. X}
  {\bfseries 5} (2015) 041025}.

\bibitem{wigner1951}
E.~Wigner, \emph{On the statistical distribution of the widths and spacings of
  nuclear resonance levels},
  \href{https://doi.org/10.1017/S0305004100027237}{\emph{Math. Proc. Cam. Phil.
  Soc.} {\bfseries 49} (1951) 790}.

\bibitem{mehta1960density}
M.L.~Mehta and M.~Gaudin, \emph{On the density of eigenvalues of a random
  matrix}, \href{https://doi.org/10.1016/0029-5582(60)90414-4}{\emph{Nuclear
  Physics} {\bfseries 18} (1960) 420}.

\bibitem{dyson1962statistical}
F.J.~Dyson, \emph{{Statistical theory of the energy levels of complex systems.
  I}}, \href{https://doi.org/10.1063/1.1703773}{\emph{J. Math. Phys.}
  {\bfseries 3} (1962) 140}.

\bibitem{dyson1962statisticalII}
F.J.~Dyson, \emph{Statistical theory of the energy levels of complex systems.
  ii}, \href{https://doi.org/10.1063/1.1702774}{\emph{Journal of Mathematical
  Physics} {\bfseries 3} (1962) 157}.

\bibitem{dyson1962statisticalIII}
F.J.~Dyson, \emph{Statistical theory of the energy levels of complex systems.
  iii}, \href{https://doi.org/10.1063/1.1703775}{\emph{Journal of Mathematical
  Physics} {\bfseries 3} (1962) 166}.

\bibitem{dyson1963statisticalI}
F.J.~Dyson and M.L.~Mehta, \emph{Statistical theory of the energy levels of
  complex systems. iv}, \href{https://doi.org/10.1063/1.1704008}{\emph{Journal
  of Mathematical Physics} {\bfseries 4} (1963) 701}.

\bibitem{mehta1963statisticalII}
M.L.~Mehta and F.J.~Dyson, \emph{Statistical theory of the energy levels of
  complex systems. v}, \href{https://doi.org/10.1063/1.1704009}{\emph{Journal
  of Mathematical Physics} {\bfseries 4} (1963) 713}.

\bibitem{dyson1962threefold}
F.J.~Dyson, \emph{The threefold way. algebraic structure of symmetry groups and
  ensembles in quantum mechanics}, {\emph{Journal of Mathematical Physics}
  {\bfseries 3} (1962) 1199}.

\bibitem{guhr1998}
T.~Guhr, A.~Muller-Groeling and H.A.~Weidenmuller, \emph{{Random matrix
  theories in quantum physics: Common concepts}},
  \href{https://doi.org/10.1016/S0370-1573(97)00088-4}{\emph{Phys. Rept.}
  {\bfseries 299} (1998) 189}
  [\href{https://arxiv.org/abs/cond-mat/9707301}{{\ttfamily
  cond-mat/9707301}}].

\bibitem{ginibre1965}
J.~Ginibre, \emph{Statistical ensembles of complex, quaternion, and real
  matrices}, \href{https://doi.org/10.1063/1.1704292}{\emph{Journal of
  Mathematical Physics} {\bfseries 6} (1965) 440}.

\bibitem{Verbaarschot1984a}
J.J.M.~Verbaarschot, H.A.~Weidenm\"uller and M.R.~Zirnbauer, \emph{{Grassmann
  integration and the theory of compound-nucleus reactions}},
  \href{https://doi.org/10.1016/0370-2693(84)90402-7}{\emph{Phys. Lett. B}
  {\bfseries 149} (1984) 263}.

\bibitem{Verbaarschot:1985jn}
J.J.M.~Verbaarschot, H.A.~Weidenmuller and M.R.~Zirnbauer, \emph{{Grassmann
  Integration in Stochastic Quantum Physics: The Case of Compound Nucleus
  Scattering}}, \href{https://doi.org/10.1016/0370-1573(85)90070-5}{\emph{Phys.
  Rept.} {\bfseries 129} (1985) 367}.

\bibitem{sommers1999s}
H.-J.~Sommers, Y.V.~Fyodorov and M.~Titov, \emph{S-matrix poles for chaotic
  quantum systems as eigenvalues of complex symmetric random matrices: from
  isolated to overlapping resonances},
  \href{https://doi.org/10.1088/0305-4470/32/5/003}{\emph{Journal of Physics A:
  Mathematical and General} {\bfseries 32} (1999) L77–L85}
  [\href{https://arxiv.org/abs/chao-dyn/9807015}{{\ttfamily
  chao-dyn/9807015}}].

\bibitem{hatano1996localization}
N.~Hatano and D.R.~Nelson, \emph{Localization transitions in non-hermitian
  quantum mechanics},
  \href{https://doi.org/10.1103/physrevlett.77.570}{\emph{Physical Review
  Letters} {\bfseries 77} (1996) 570–573}
  [\href{https://arxiv.org/abs/cond-mat/9603165}{{\ttfamily
  cond-mat/9603165}}].

\bibitem{efetov1997directed}
K.B.~Efetov, \emph{Directed quantum chaos},
  \href{https://doi.org/10.1103/physrevlett.79.491}{\emph{Physical Review
  Letters} {\bfseries 79} (1997) 491–494}
  [\href{https://arxiv.org/abs/cond-mat/9702091}{{\ttfamily
  cond-mat/9702091}}].

\bibitem{brouwer1998delocalization}
P.W.~Brouwer, C.~Mudry, B.D.~Simons and A.~Altland, \emph{Delocalization in
  coupled one-dimensional chains},
  \href{https://doi.org/10.1103/physrevlett.81.862}{\emph{Physical Review
  Letters} {\bfseries 81} (1998) 862–865}
  [\href{https://arxiv.org/abs/cond-mat/9807189}{{\ttfamily
  cond-mat/9807189}}].

\bibitem{fyodorov1997}
Y.V.~Fyodorov, B.A.~Khoruzhenko and H.-J.~Sommers, \emph{Almost hermitian
  random matrices: Crossover from wigner-dyson to ginibre eigenvalue
  statistics}, \href{https://doi.org/10.1103/physrevlett.79.557}{\emph{Physical
  Review Letters} {\bfseries 79} (1997) 557–560}.

\bibitem{akemann2019universal}
G.~Akemann, M.~Kieburg, A.~Mielke and T.~Prosen, \emph{Universal signature from
  integrability to chaos in dissipative open quantum systems},
  \href{https://doi.org/10.1103/physrevlett.123.254101}{\emph{Physical Review
  Letters} {\bfseries 123} (2019) }
  [\href{https://arxiv.org/abs/arXiv:1910.03520}{{\ttfamily
  arXiv:1910.03520}}].

\bibitem{li2021spectral}
J.~Li, T.~Prosen and A.~Chan, \emph{{Spectral Statistics of Non-Hermitian
  Matrices and Dissipative Quantum Chaos}},
  \href{https://doi.org/10.1103/PhysRevLett.127.170602}{\emph{Phys. Rev. Lett.}
  {\bfseries 127} (2021) 170602}
  [\href{https://arxiv.org/abs/2103.05001}{{\ttfamily 2103.05001}}].

\bibitem{sa2021lindbladian}
L.~S\'a, P.~Ribeiro and T.~Prosen, \emph{{Lindbladian dissipation of
  strongly-correlated quantum matter}},
  \href{https://arxiv.org/abs/2112.12109}{{\ttfamily 2112.12109}}.

\bibitem{stephanov1996}
M.A.~Stephanov, \emph{{Random matrix model of QCD at finite density and the
  nature of the quenched limit}},
  \href{https://doi.org/10.1103/PhysRevLett.76.4472}{\emph{Phys. Rev. Lett.}
  {\bfseries 76} (1996) 4472}
  [\href{https://arxiv.org/abs/hep-lat/9604003}{{\ttfamily hep-lat/9604003}}].

\bibitem{Janik:1996va}
R.A.~Janik, M.A.~Nowak, G.~Papp and I.~Zahed, \emph{{Brezin-Zee universality:
  Why quenched QCD in matter is subtle?}},
  \href{https://doi.org/10.1103/PhysRevLett.77.4876}{\emph{Phys. Rev. Lett.}
  {\bfseries 77} (1996) 4876}
  [\href{https://arxiv.org/abs/hep-ph/9606329}{{\ttfamily hep-ph/9606329}}].

\bibitem{Verbaarschot:2000dy}
J.J.M.~Verbaarschot and T.~Wettig, \emph{{Random matrix theory and chiral
  symmetry in QCD}},
  \href{https://doi.org/10.1146/annurev.nucl.50.1.343}{\emph{Ann. Rev. Nucl.
  Part. Sci.} {\bfseries 50} (2000) 343}
  [\href{https://arxiv.org/abs/hep-ph/0003017}{{\ttfamily hep-ph/0003017}}].

\bibitem{Osborn:2004rf}
J.C.~Osborn, \emph{{Universal results from an alternate random matrix model for
  QCD with a baryon chemical potential}},
  \href{https://doi.org/10.1103/PhysRevLett.93.222001}{\emph{Phys. Rev. Lett.}
  {\bfseries 93} (2004) 222001}
  [\href{https://arxiv.org/abs/hep-th/0403131}{{\ttfamily hep-th/0403131}}].

\bibitem{Akemann:2004dr}
G.~Akemann, J.C.~Osborn, K.~Splittorff and J.J.M.~Verbaarschot,
  \emph{{Unquenched QCD Dirac operator spectra at nonzero baryon chemical
  potential}},
  \href{https://doi.org/10.1016/j.nuclphysb.2005.01.018}{\emph{Nucl. Phys. B}
  {\bfseries 712} (2005) 287}
  [\href{https://arxiv.org/abs/hep-th/0411030}{{\ttfamily hep-th/0411030}}].

\bibitem{Osborn:2005ss}
J.C.~Osborn, K.~Splittorff and J.J.M.~Verbaarschot, \emph{{Chiral symmetry
  breaking and the Dirac spectrum at nonzero chemical potential}},
  \href{https://doi.org/10.1103/PhysRevLett.94.202001}{\emph{Phys. Rev. Lett.}
  {\bfseries 94} (2005) 202001}
  [\href{https://arxiv.org/abs/hep-th/0501210}{{\ttfamily hep-th/0501210}}].

\bibitem{Kanazawa:2009en}
T.~Kanazawa, T.~Wettig and N.~Yamamoto, \emph{{Chiral random matrix theory for
  two-color QCD at high density}},
  \href{https://doi.org/10.1103/PhysRevD.81.081701}{\emph{Phys. Rev. D}
  {\bfseries 81} (2010) 081701}
  [\href{https://arxiv.org/abs/0912.4999}{{\ttfamily 0912.4999}}].

\bibitem{kanazawa2021new}
T.~Kanazawa and T.~Wettig, \emph{{New universality classes of the non-Hermitian
  Dirac operator in QCD-like theories}},
  \href{https://doi.org/10.1103/PhysRevD.104.014509}{\emph{Phys. Rev. D}
  {\bfseries 104} (2021) 014509}
  [\href{https://arxiv.org/abs/2104.05846}{{\ttfamily 2104.05846}}].

\bibitem{kanazawa2021complex}
T.~Kanazawa and T.~Wettig, \emph{{Complex spacing ratios of the non-Hermitian
  Dirac operator in universality classes AI$^\dagger$ and AII$^\dagger$}},  in
  \emph{{38th International Symposium on Lattice Field Theory}}, 11, 2021
  [\href{https://arxiv.org/abs/2111.04573}{{\ttfamily 2111.04573}}].

\bibitem{bender1998}
C.M.~Bender and S.~Boettcher, \emph{{Real spectra in nonHermitian Hamiltonians
  having PT symmetry}},
  \href{https://doi.org/10.1103/PhysRevLett.80.5243}{\emph{Phys. Rev. Lett.}
  {\bfseries 80} (1998) 5243}
  [\href{https://arxiv.org/abs/physics/9712001}{{\ttfamily physics/9712001}}].

\bibitem{halasz:1997fc}
A.M.~Halasz, J.C.~Osborn and J.J.M.~Verbaarschot, \emph{{Random matrix triality
  at nonzero chemical potential}},
  \href{https://doi.org/10.1103/PhysRevD.56.7059}{\emph{Phys. Rev. D}
  {\bfseries 56} (1997) 7059}
  [\href{https://arxiv.org/abs/hep-lat/9704007}{{\ttfamily hep-lat/9704007}}].

\bibitem{bernard2002classification}
D.~Bernard and A.~LeClair, \emph{A classification of non-hermitian random
  matrices},
  \href{https://doi.org/10.1007/978-94-010-0514-2_19}{\emph{Statistical Field
  Theories} (2002) 207–214}
  [\href{https://arxiv.org/abs/cond-mat/0110649}{{\ttfamily
  cond-mat/0110649}}].

\bibitem{Magnea:2007yk}
U.~Magnea, \emph{{Random matrices beyond the Cartan classification}},
  \href{https://doi.org/10.1088/1751-8113/41/4/045203}{\emph{J. Phys. A}
  {\bfseries 41} (2008) 045203}
  [\href{https://arxiv.org/abs/0707.0418}{{\ttfamily 0707.0418}}].

\bibitem{kawabata2019symmetry}
K.~Kawabata, K.~Shiozaki, M.~Ueda and M.~Sato, \emph{{Symmetry and Topology in
  Non-Hermitian Physics}},
  \href{https://doi.org/10.1103/PhysRevX.9.041015}{\emph{Phys. Rev. X}
  {\bfseries 9} (2019) 041015}
  [\href{https://arxiv.org/abs/1812.09133}{{\ttfamily 1812.09133}}].

\bibitem{Garcia-Garcia:2021rle}
A.M.~Garc\'\i{}a-Garc\'\i{}a, L.~S\'a and J.J.M.~Verbaarschot, \emph{{Symmetry
  classification and universality in non-Hermitian many-body quantum chaos by
  the Sachdev-Ye-Kitaev model}},
  \href{https://arxiv.org/abs/2110.03444}{{\ttfamily 2110.03444}}.

\bibitem{Verbaarschot:1994qf}
J.J.M.~Verbaarschot, \emph{{The Spectrum of the QCD Dirac operator and chiral
  random matrix theory: The Threefold way}},
  \href{https://doi.org/10.1103/PhysRevLett.72.2531}{\emph{Phys. Rev. Lett.}
  {\bfseries 72} (1994) 2531}
  [\href{https://arxiv.org/abs/hep-th/9401059}{{\ttfamily hep-th/9401059}}].

\bibitem{Altland:1997zz}
A.~Altland and M.R.~Zirnbauer, \emph{{Nonstandard symmetry classes in
  mesoscopic normal-superconducting hybrid structures}},
  \href{https://doi.org/10.1103/PhysRevB.55.1142}{\emph{Phys. Rev. B}
  {\bfseries 55} (1997) 1142}
  [\href{https://arxiv.org/abs/cond-mat/9602137}{{\ttfamily
  cond-mat/9602137}}].

\bibitem{ashida2020non}
Y.~Ashida, Z.~Gong and M.~Ueda, \emph{{Non-Hermitian physics}},
  \href{https://doi.org/10.1080/00018732.2021.1876991}{\emph{Adv. Phys.}
  {\bfseries 69} (2021) 249}
  [\href{https://arxiv.org/abs/2006.01837}{{\ttfamily 2006.01837}}].

\bibitem{maldacena2018}
J.~Maldacena and X.-L.~Qi, \emph{{Eternal traversable wormhole}},
  \href{https://arxiv.org/abs/1804.00491}{{\ttfamily 1804.00491}}.

\bibitem{Garcia-Garcia:2019poj}
A.M.~Garc\'\i{}a-Garc\'\i{}a, T.~Nosaka, D.~Rosa and J.J.M.~Verbaarschot,
  \emph{{Quantum chaos transition in a two-site Sachdev-Ye-Kitaev model dual to
  an eternal traversable wormhole}},
  \href{https://doi.org/10.1103/PhysRevD.100.026002}{\emph{Phys. Rev. D}
  {\bfseries 100} (2019) 026002}
  [\href{https://arxiv.org/abs/1901.06031}{{\ttfamily 1901.06031}}].

\bibitem{Garcia-Garcia:2020ttf}
A.M.~Garc\'\i{}a-Garc\'\i{}a and V.~Godet, \emph{{Euclidean wormhole in the
  Sachdev-Ye-Kitaev model}},
  \href{https://doi.org/10.1103/PhysRevD.103.046014}{\emph{Phys. Rev. D}
  {\bfseries 103} (2021) 046014}
  [\href{https://arxiv.org/abs/2010.11633}{{\ttfamily 2010.11633}}].

\bibitem{Garcia-Garcia:2021elz}
A.M.~Garc\'\i{}a-Garc\'\i{}a, Y.~Jia, D.~Rosa and J.J.M.~Verbaarschot,
  \emph{{Replica Symmetry Breaking and Phase Transitions in a PT Symmetric
  Sachdev-Ye-Kitaev Model}},
  \href{https://arxiv.org/abs/2102.06630}{{\ttfamily 2102.06630}}.

\bibitem{Garcia-Garcia:2022}
A.M.~Garc\'\i{}a-Garc\'\i{}a, Y.~Jia, D.~Rosa and J.J.M.~Verbaarschot,
  \emph{{Replica Symmetry Breaking in Random Non-Hermitian Systems}},  2022.

\bibitem{dyson1949s}
F.J.~Dyson, \emph{{The S matrix in quantum electrodynamics}},
  \href{https://doi.org/10.1103/PhysRev.75.1736}{\emph{Phys. Rev.} {\bfseries
  75} (1949) 1736}.

\bibitem{schwinger1951green}
J.S.~Schwinger, \emph{{On the Green's functions of quantized fields. 1.}},
  \href{https://doi.org/10.1073/pnas.37.7.452}{\emph{Proc. Nat. Acad. Sci.}
  {\bfseries 37} (1951) 452}.

\bibitem{Maldacena:2015waa}
J.~Maldacena, S.H.~Shenker and D.~Stanford, \emph{{A bound on chaos}},
  \href{https://doi.org/10.1007/JHEP08(2016)106}{\emph{JHEP} {\bfseries 08}
  (2016) 106} [\href{https://arxiv.org/abs/1503.01409}{{\ttfamily
  1503.01409}}].

\bibitem{Cotler:2016fpe}
J.S.~Cotler, G.~Gur-Ari, M.~Hanada, J.~Polchinski, P.~Saad, S.H.~Shenker
  et~al., \emph{{Black Holes and Random Matrices}},
  \href{https://doi.org/10.1007/JHEP05(2017)118}{\emph{JHEP} {\bfseries 05}
  (2017) 118} [\href{https://arxiv.org/abs/1611.04650}{{\ttfamily
  1611.04650}}].

\bibitem{edwards1971statistical}
S.~Edwards, \emph{The statistical mechanics of rubbers},
  \href{https://doi.org/10.1007/978-1-4757-6210-5_5}{\emph{Polymer Networks}
  (1971) 83}.

\bibitem{sherrington1972}
D.~Sherrington and S.~Kirkpatrick, \emph{Solvable model of a spin-glass},
  \href{https://doi.org/10.1103/PhysRevLett.35.1792}{\emph{Phys. Rev. Lett.}
  {\bfseries 35} (1975) 1792}.

\bibitem{verbaarschot1985}
J.J.M.~Verbaarschot and M.R.~Zirnbauer, \emph{Critique of the replica trick},
  \href{https://doi.org/10.1088/0305-4470/18/7/018}{\emph{Journal of Physics A:
  Mathematical and General} {\bfseries 18} (1985) 1093}.

\bibitem{zirnbauer1999critique}
M.R.~Zirnbauer, \emph{Another critique of the replica trick},
  \href{https://arxiv.org/abs/cond-mat/9903338}{{\ttfamily cond-mat/9903338}}.

\bibitem{Barbour:1986jf}
I.~Barbour, N.-E.~Behilil, E.~Dagotto, F.~Karsch, A.~Moreo, M.~Stone et~al.,
  \emph{{Problems with Finite Density Simulations of Lattice QCD}},
  \href{https://doi.org/10.1016/0550-3213(86)90601-2}{\emph{Nucl. Phys. B}
  {\bfseries 275} (1986) 296}.

\bibitem{parisi1979}
G.~Parisi, \emph{Toward a mean field theory for spin glasses},
  \href{https://doi.org/10.1016/0375-9601(79)90708-4}{\emph{Physics Letters A}
  {\bfseries 73} (1979) 203}.

\bibitem{girko2012theory}
V.L.~Girko, \emph{Theory of random determinants}, vol.~45, Springer Science \&
  Business Media (2012).

\bibitem{mezard1999}
A.~Kamenev and M.~Mézard, \emph{Wigner-dyson statistics from the replica
  method}, \href{https://doi.org/10.1088/0305-4470/32/24/304}{\emph{Journal of
  Physics A: Mathematical and General} {\bfseries 32} (1999) 4373–4388}.

\bibitem{nishigaki2002a}
S.M.~Nishigaki and A.~Kamenev, \emph{Replica treatment of non-hermitian
  disordered hamiltonians},
  \href{https://doi.org/10.1088/0305-4470/35/21/307}{\emph{Journal of Physics
  A: Mathematical and General} {\bfseries 35} (2002) 4571–4590}.

\bibitem{kanzieper2002}
E.~Kanzieper, \emph{{Replica field theories, Painleve transcendents and exact
  correlation functions}},
  \href{https://doi.org/10.1103/PhysRevLett.89.250201}{\emph{Phys. Rev. Lett.}
  {\bfseries 89} (2002) 250201}
  [\href{https://arxiv.org/abs/cond-mat/0207745}{{\ttfamily
  cond-mat/0207745}}].

\bibitem{splittorff:2003cu}
K.~Splittorff and J.J.M.~Verbaarschot, \emph{{Factorization of correlation
  functions and the replica limit of the Toda lattice equation}},
  \href{https://doi.org/10.1016/j.nuclphysb.2004.01.031}{\emph{Nucl. Phys. B}
  {\bfseries 683} (2004) 467}
  [\href{https://arxiv.org/abs/hep-th/0310271}{{\ttfamily hep-th/0310271}}].

\bibitem{sedrakyan2005toda}
T.A.~Sedrakyan, \emph{{Toda lattice representation for random matrix model with
  logarithmic confinement}},
  \href{https://doi.org/10.1016/j.nuclphysb.2005.09.020}{\emph{Nucl. Phys. B}
  {\bfseries 729} (2005) 526}
  [\href{https://arxiv.org/abs/cond-mat/0506373}{{\ttfamily
  cond-mat/0506373}}].

\bibitem{arefeva2018}
I.~Aref'eva, M.~Khramtsov, M.~Tikhanovskaya and I.~Volovich,
  \emph{{Replica-nondiagonal solutions in the SYK model}},
  \href{https://doi.org/10.1007/JHEP07(2019)113}{\emph{JHEP} {\bfseries 07}
  (2019) 113} [\href{https://arxiv.org/abs/1811.04831}{{\ttfamily
  1811.04831}}].

\bibitem{wang2018}
H.~Wang, D.~Bagrets, A.L.~Chudnovskiy and A.~Kamenev, \emph{{On the replica
  structure of Sachdev-Ye-Kitaev model}},
  \href{https://doi.org/10.1007/JHEP09(2019)057}{\emph{JHEP} {\bfseries 09}
  (2019) 057} [\href{https://arxiv.org/abs/1812.02666}{{\ttfamily
  1812.02666}}].

\bibitem{feinberg1997non}
J.~Feinberg and A.~Zee, \emph{{NonGaussian nonHermitian random matrix theory:
  Phase transition and addition formalism}},
  \href{https://doi.org/10.1016/S0550-3213(97)00419-7}{\emph{Nucl. Phys. B}
  {\bfseries 501} (1997) 643}
  [\href{https://arxiv.org/abs/cond-mat/9704191}{{\ttfamily
  cond-mat/9704191}}].

\bibitem{Janik:1996xm}
R.A.~Janik, M.A.~Nowak, G.~Papp and I.~Zahed, \emph{{NonHermitian random matrix
  models. 1.}},
  \href{https://doi.org/10.1016/S0550-3213(97)00418-5}{\emph{Nucl. Phys. B}
  {\bfseries 501} (1997) 603}
  [\href{https://arxiv.org/abs/cond-mat/9612240}{{\ttfamily
  cond-mat/9612240}}].

\bibitem{Saad:2019lba}
P.~Saad, S.H.~Shenker and D.~Stanford, \emph{{JT gravity as a matrix
  integral}},  \href{https://arxiv.org/abs/1903.11115}{{\ttfamily 1903.11115}}.

\bibitem{Altland:2021rqn}
A.~Altland, D.~Bagrets, P.~Nayak, J.~Sonner and M.~Vielma, \emph{{From operator
  statistics to wormholes}},
  \href{https://doi.org/10.1103/PhysRevResearch.3.033259}{\emph{Phys. Rev.
  Res.} {\bfseries 3} (2021) 033259}
  [\href{https://arxiv.org/abs/2105.12129}{{\ttfamily 2105.12129}}].

\bibitem{garcia2018chaotic}
A.M.~García-García, B.~Loureiro, A.~Romero-Bermúdez and M.~Tezuka,
  \emph{Chaotic-integrable transition in the sachdev-ye-kitaev model},
  \href{https://doi.org/10.1103/physrevlett.120.241603}{\emph{Physical Review
  Letters} {\bfseries 120} (2018) }
  [\href{https://arxiv.org/abs/1707.02197}{{\ttfamily 1707.02197}}].

\bibitem{hastings2001}
M.~Hastings, \emph{Eigenvalue distribution in the self-dual non-hermitian
  ensemble}, \href{https://doi.org/10.1023/A:1010356821471}{\emph{Journal of
  Statistical Physics} {\bfseries 103} (2001) 903}
  [\href{https://arxiv.org/abs/cond-mat/9909234}{{\ttfamily
  cond-mat/9909234}}].

\bibitem{hamazaki2020}
R.~Hamazaki, K.~Kawabata, N.~Kura and M.~Ueda, \emph{Universality classes of
  non-hermitian random matrices},
  \href{https://doi.org/10.1103/PhysRevResearch.2.023286}{\emph{Phys. Rev.
  Research} {\bfseries 2} (2020) 023286}
  [\href{https://arxiv.org/abs/1904.13082}{{\ttfamily 1904.13082}}].

\bibitem{akemann2022spacing}
G.~Akemann, A.~Mielke and P.~Päßler, \emph{Spacing distribution in the 2d
  coulomb gas: Surmise and symmetry classes of non-hermitian random matrices at
  non-integer $\beta$},  2022.

\bibitem{Kanazawa:2014qma}
T.~Kanazawa and Y.~Tanizaki, \emph{{Structure of Lefschetz thimbles in simple
  fermionic systems}},
  \href{https://doi.org/10.1007/JHEP03(2015)044}{\emph{JHEP} {\bfseries 03}
  (2015) 044} [\href{https://arxiv.org/abs/1412.2802}{{\ttfamily 1412.2802}}].

\bibitem{Tanizaki:2015gpl}
Y.~Tanizaki, \emph{{Study on sign problem via Lefschetz-thimble path
  integral}}, Ph.D. thesis, Tokyo U., 12, 2015.
\newblock 10.15083/00073296.

\bibitem{Liao:2021ofk}
Y.~Liao and V.~Galitski, \emph{{Emergence of many-body quantum chaos via
  spontaneous breaking of unitarity}},
  \href{https://arxiv.org/abs/2104.05721}{{\ttfamily 2104.05721}}.

\bibitem{Banks:1979yr}
T.~Banks and A.~Casher, \emph{{Chiral Symmetry Breaking in Confining
  Theories}}, \href{https://doi.org/10.1016/0550-3213(80)90255-2}{\emph{Nucl.
  Phys. B} {\bfseries 169} (1980) 103}.

\bibitem{Jackson:1995nf}
A.D.~Jackson and J.J.M.~Verbaarschot, \emph{{A Random matrix model for chiral
  symmetry breaking}},
  \href{https://doi.org/10.1103/PhysRevD.53.7223}{\emph{Phys. Rev. D}
  {\bfseries 53} (1996) 7223}
  [\href{https://arxiv.org/abs/hep-ph/9509324}{{\ttfamily hep-ph/9509324}}].

\bibitem{Jackson:1996xt}
A.D.~Jackson, M.K.~Sener and J.J.M.~Verbaarschot, \emph{{Universality near zero
  virtuality}}, \href{https://doi.org/10.1016/0550-3213(96)00397-5}{\emph{Nucl.
  Phys. B} {\bfseries 479} (1996) 707}
  [\href{https://arxiv.org/abs/hep-ph/9602225}{{\ttfamily hep-ph/9602225}}].

\bibitem{Sachdev:2001}
A.~Georges, O.~Parcollet and S.~Sachdev, \emph{Quantum fluctuations of a nearly
  critical heisenberg spin glass},
  \href{https://doi.org/10.1103/physrevb.63.134406}{\emph{Physical Review B}
  {\bfseries 63} (2001) }.

\end{thebibliography}\endgroup
 
\end{document}